\documentclass[times,twocolumn,final]{elsarticle}

\usepackage{preprint}
\usepackage{framed,multirow}
\usepackage{amssymb}
\usepackage{latexsym}
\usepackage{url}
\usepackage{xcolor}
\definecolor{newcolor}{rgb}{.8,.349,.1}
\usepackage[
    colorlinks=true,
    linkcolor=blue,
    filecolor=blue,      
    urlcolor=blue]
{hyperref}
\usepackage{nicefrac}
\usepackage{amsmath}
\usepackage{booktabs}
\usepackage{wrapfig}
\usepackage{subcaption} 
\usepackage[ruled,vlined]{algorithm2e}
\usepackage[switch,pagewise]{lineno} 
\SetAlFnt{\footnotesize}
\SetAlCapFnt{\small}
\SetAlCapNameFnt{\small}

    \newcommand{\rem}[1]{}

    \newcommand{\etal}{{et al.~}}
    
    \newcommand{\step}[1]{{\color{black}#1}} 
    \newcommand{\stepR}[1]{{\color{black}#1}} 

    \newcommand{\tod}[1]{\overline{#1}}
    
    \newcommand{\injR}{{ir}}

    \newcommand{\cladf}{\mathcal{F}} 
    
    \newcommand{\NRMSE}{\text{\textit{NRMSE}}}
    
\newcommand\bmat[1]{\begin{bmatrix}#1\end{bmatrix}} 

\begin{document}
\nolinenumbers

\verso{Accepted for Publication}

\begin{frontmatter}

\title{Design and Fabrication of Multi-Patch Elastic Geodesic Grid Structures}

\author[1]{Stefan {Pillwein}\corref{cor1}}
\cortext[cor1]{Corresponding author:}
\emailauthor{stefan.pillwein@tuwien.ac.at}{Stefan Pillwein}
\author[1]{Johanna {K\"ubert}}
\author[1,2]{Florian {Rist}}
\author[1,3]{Przemyslaw {Musialski}} 

\address[1]{Technische Universit\"at Wien (TU Wien)}
\address[2]{King Abdullah University of Science and Technology (KAUST)}
\address[3]{New Jersey Institute of Technology (NJIT)}

\accepted{June 2, 2021}

\begin{abstract}
Elastic geodesic grids (EGG) are lightweight structures that can be deployed to approximate designer-provided free-form surfaces. Initially, the grids are perfectly flat, during deployment, a curved shape emerges, as grid elements bend and twist. Their layout is based on networks of geodesic curves and is found geometrically. Encoded in the planar grids is the intrinsic shape of the design surface. Such structures may serve purposes like free-form sub-structures, panels, sun and rain protectors, pavilions, etc. However, so far the EGG have only been investigated using a generic set of design surfaces and small-scale desktop models. Some limitations become apparent when considering more sophisticated design surfaces, like from free-form architecture. Due to characteristics like high local curvature or non-geodesic boundaries, they may be captured only poorly by a single EGG. 

We show how decomposing such surfaces into smaller patches serves as an effective strategy to tackle these problems. We furthermore show that elastic geodesic grids are in fact well suited for this approach. Finally, we present a showcase model of some meters in size and discuss practical aspects concerning fabrication, size, and easy deployment. 
\end{abstract}

\begin{keyword}
\KWD Computer graphics and computational geometry, Computer graphics; computational geometry
\end{keyword}

\end{frontmatter}

\nolinenumbers

\section{Introduction}\label{sec:introduction}
\begin{figure}
	\centering
	\includegraphics[width=\columnwidth]{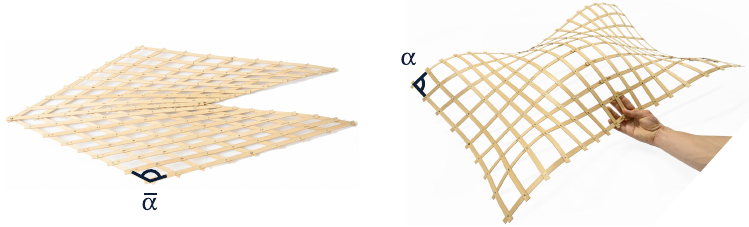}
	\caption{An elastic geodesic grid in the planar and the deployed configuration, the deployment is based on $ \tod{\alpha} \rightarrow \alpha$.}
	\label{fig:pla_dep}
    \end{figure}

    In design and architecture, curved structures are aesthetically pleasing and functional, however, they are not always easy to produce. 
    The curved shape of the structure in Figure \ref{fig:pla_dep} was generated by deploying a planar grid of initially straight but elastic elements fabricated as wooden lamellae. Such grids, referred to as Elastic Geodesic Grids (EGG) by Pillwein \etal\cite{Pillwein2020}, can be deployed from a planar state to a spatial state that approximates a given design surface. 
    \step{The form-finding approach of EGG encodes the shape of the surface into the grid and produces planar grids with nonparallel members.  
    This is the key factor for their deployment: 
    The grids are rigid in the plane, but may easily buckle out of the plane and expand to their spatial configuration.
    This transformation can be performed simply by expanding the compact planar setup similar to a scissor-like mechanism. }

    This deployment approach makes the creation of doubly-curved shapes quick and material efficient, and the lightness and easy assembly make the grids applicable for mobile purposes. In addition, their fabrication does not need complicated techniques or advanced materials. We present a showcase model, which was built using simple resources, like plywood, screws, washers, and nuts. If advanced tools are not available, it could even be built using a tape measure, a drill, and a saw.
    Lightweight grid structures are particularly suitable for applications where large spans and low weight are required, which makes them potential candidates for architectural purposes. 
    
\begin{figure}
	\centering
	\includegraphics[trim=0 28 0 58,clip,width=0.68\columnwidth]{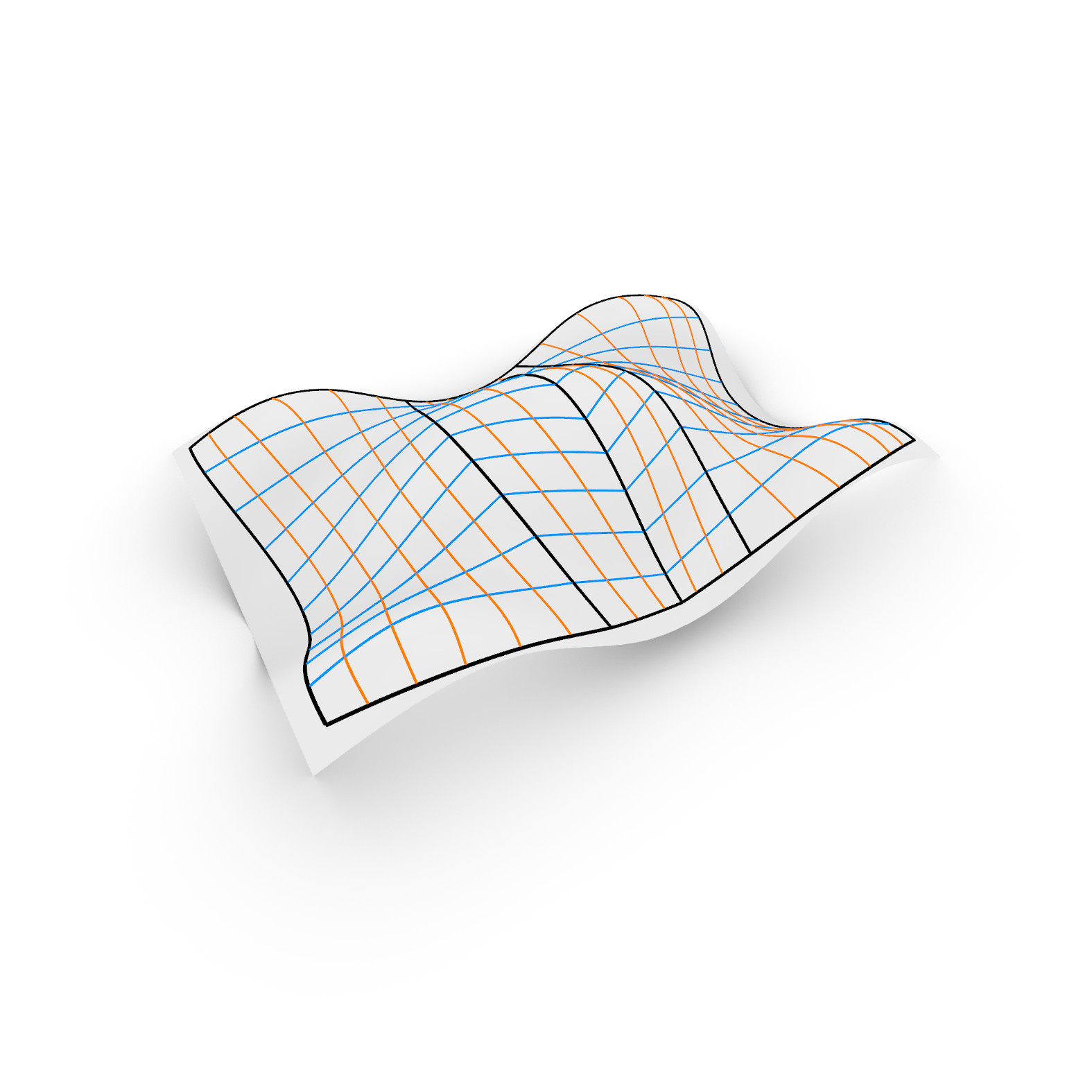}
	\vspace{-30pt}
	\caption{A surface is split into multiple patches, boundaries are indicated in black, grid members in orange and blue. Blue members share start and endpoints.}
	\label{fig:patch_boundaries}
    \end{figure}

    \step{A variety of shapes can be approximated with the EGG approach: they can approximate elliptic and hyperbolic surfaces (i.e., they may have positive as well as negative Gaussian curvature $K$), and multiple changes in curvature are also possible. 
    Although the range of feasible surfaces is wide, the approach has a drawback for designers: the part of the surface that can be used for an elastic grid needs to be framed by a geodesic quadrilateral, which means the rest of the surface will be cut off.
    
    Another set of problems comes with high Gaussian curvature $K$: First, if the peaks are too high, the strips on the surface (cf. Figure \ref{fig:pla_dep}, right) will become very long, and at some point, it becomes impossible to fit them into the planar configuration. \stepR{Second, high-$K$ areas may cause geodesics to be non-unique, which makes the surface unrepresentable using the EGG approach. Previously these issues were solved by smoothing the surface \cite{Pillwein2020} until these problems vanish. This, however, is not desirable as it alters the target surface and its characteristics.}
    
    These problems can also be dealt with by decomposing the input surface into multiple patches. This approach omits the smoothing procedure and ensures better coverage of the design surface. 
    Along the common boundary of two adjacent patches, the smoothness of the geodesic curves is broken, but the position and thus $C^0$-continuity is preserved. The boundaries of patches still need to be geodesic curves, but the placement and size of the patches can be chosen following the geometric features of the target shape. This splitting approach fits well with the idea of elastic geodesic grids, and a byproduct is a parameterization of the patch with geodesic curves.
    
    Computing parameterizations for all patches before defining the members of the grid, allows an aesthetically pleasing and even distribution of members over the entire group of patches on the entire design surface.
    Figure \ref{fig:patch_boundaries} shows how such a parameterization can be used to propagate members across multiple patches.
    }

    \step{
    In this paper, we extend the recently presented work of Pillwein \etal\cite{Pillwein2020,Pillwein2020a}, in particular, our goal is to analyze how elastic geodesic grid models can be designed on challenging input surfaces that require multiple patches. We provide the following contributions:
    \begin{itemize}
        \item The geometric background of efficiently decomposing large and challenging input surfaces is discussed in detail. Such surfaces can have holes and non-convex boundaries. Practical rules for this process are presented and applied to three examples of free-form architecture.
        \item The propagation of grid members across patches and their even distribution is introduced. 
        \item A showcase model and its fabrication are presented. Practical aspects like material, size, and strength are discussed.
    \end{itemize}
    }

    In Section \ref{sec:related_work} we give an overview of the related work, 
    in Section \ref{sec:egg} we review the EGG approach, 
    in Section \ref{sec:modularity} we examine how surfaces can be split effectively, and 
    in Section \ref{sec:challenges} we discuss fabrication and design challenges. A showcase model, the fabrication process, and three design studies are presented in Section \ref{sec:pavilion}. Eventually, in Section \ref{sec:discussion} we feature a discussion of advantages and limitations.

    \section{Related Work}\label{sec:related_work}
    \paragraph{Active-Bending Paradigm}
    In the computer science, architecture, and engineering communities, the \textit{active bending} paradigm \cite{Lienhard2013} and easy-to-use computational methods have sparked a new wave of interest in elastic structures that are deformed to create curved shapes.
    Until recent advances in computing elastic structures fast and physically accurate, they could only be form-found empirically \cite{gengnagel2013active}. 
  
    \paragraph{Gridshells and Deployment-Approaches}
    A lot of research is currently being carried out on gridshell-structures that can be deployed. They can be classified based on their deployment mechanism: inscribed in a grid that is deployed \cite{Panetta2019,X_shells_pavilion,Soriano2019,Pillwein2020} or by other external mechanisms like inflatable air cushions \cite{pneumatic}.
    We are interested in the first case. The design approaches of inscribing the deployment mechanism into the grid, however, differ a lot: 
    
    To create an X-Shell \cite{Panetta2019}, a planar grid layout is designed and actuated with a physical simulator. The grid curves of the planar layout do not have to be straight. In multiple layout iterations, the designer finds a satisfactory shape by changing the planar design. This approach does not depend on a target surface. Target surfaces, however, can be approximated if a good planar initialization is provided by the designer. In this case, a grid that approximates the target surface closely can be found using shape optimization.
    The bending and twisting behavior of the rods can be controlled by using different shapes of cross-sections. This has a direct effect on the shape of the deployed X-Shell, as the paper shows.
    The practical feasibility of these structures was investigated hands-on with the construction of a pavilion of GFRP-rods, measuring 3.2 × 3.2 × 3.6 meters \cite{X_shells_pavilion}.
    
    The G-Shells approach \cite{Soriano2019} proposes to planarize a specific geodesic grid using physical simulation inside an evolutionary multi-objective solver. This way, a geometric error is introduced, so the flat grid cannot be deployed to match the geodesic grid perfectly, but it can come close, creating beautiful shapes. G-Shells look similar to elastic geodesic grids, as they are also built from thin lamellae.
    
    The EGG approach \cite{Pillwein2020} takes a design surface as input and results in a planar grid layout that can be deployed to approximate this surface. The layout is found geometrically and simultaneously on the design surface and in the plane. This makes the computation of grid layouts efficient, as neither design iterations nor physical simulations are needed for finding layouts.
    Despite the similarity of using geodesics as grid curves, in contrast to the G-Shells approach, EGG uses the concept of notches. This prevents introducing a geometric error in the grid layout, but also makes the deployment process more complicated, as sliding of members is necessary.
    However, this method does not feature arbitrary curve networks, as it is limited to geodesics as grid curves. This poses a restriction on the space of solutions of deployable grids using an inscribed deployment mechanism. Besides this limitation, design surfaces with high Gaussian curvature might need to be smoothed before a suitable grid layout is feasible.
    \step{Some practical aspects of the EGG, like reducing friction in the joints or insights on scaling, were already investigated \cite{Pillwein2020a}. The paper initially considers splitting design surfaces into multiple patches and provides first insights into it. However, in contrast to this paper, it does not explore splitting on an applied level. Also, the distribution and propagation of members across patches are not discussed.
    }

    Asymptotic curve networks were also explored as a basis for free-form structures \cite{Schling2018}. Aside from their aesthetic qualities, these structures can be assembled from initially flat parts. They naturally transform to a curved state by their internal forces. \stepR{This process induces a high twist of the members, which leads to a nonlinear geometric effect called helix-strain.
    It is typically negligible in conventional structures but needs to be taken into account for bendable lamella structures. Recently this effect was investigated for such structures specifically \cite{schikore2019}}.
    However, asymptotic curve networks can only be found on surfaces with negative Gaussian curvature.
    
    \paragraph{Other Deployable Surfaces}
    Besides the domain of architecture, there has also been extensive research in the computer graphics literature on various methods for deployable structures. 
    One more way to easily construct spatial shapes from flat sheets is by appropriately folding paper \cite{Mitani2004b,Massarwi2007a}, which is inherently related to the Japanese art of Origami \cite{Dudte2016}. 
    Other works deal with curved folds and efficient actuation of spatial objects from flat sheets \cite{Kilian2008,Kilian2017}. 
    Elastic geodesic grids are related to these approaches in terms of being deployable from a planar initial state, however, the main difference is that they are elastic and approximate doubly-curved surfaces.  

    In fact, a lot of attention has been paid to the design of doubly-curved surfaces which can be deployed from planar configurations due to the ease of fabrication. One way of achieving this goal is by using auxetic materials \cite{Konakovic2016} which can nestle to doubly-curved spatial objects, or in combination with appropriate actuation techniques, can be used to construct complex spatial objects \cite{Konakovic-Lukovic2018}. The main difference to the EGG method \cite{Pillwein2020} is that these structures do not use elastic bending to reach the spatial shape. 
    
    \stepR{Another recent approach \cite{Malomo2018a} is based on mesostructures that allow local changes in material behavior and can be connected to match target shapes. This approach has been tested on a larger scale by constructing a pavilion \cite{laccone2019flexmaps}. In contrast to EGG, mesostructures are assembled from flat, flexible panels and do not rely on a scissor-like deployment mechanism. }

    Also, the idea of storing energy in a planar configuration to approximate spatial shapes has been explored. This can be done, for instance, by using prestressed latex membranes to actuate planar structures into free-form shapes \cite{Guseinov2017}, or to predefine flexible micro-structures which deform to desired shapes if they are combined and interact \cite{Malomo2018a}. A combination of flexible rods and prestressed membranes leads to Kirchhoff-Plateau surfaces that allow easy planar fabrication and deployment \cite{Perez2017a}. 
    The EGG approach is based on the assumption that the elastic elements can bend and twist, but not stretch, and must therefore maintain the same length in the planar as well as in the spatial configuration.

    \section{Elastic Geodesic Grids (EGG)}\label{sec:egg}
    Using geodesics as grid curves is mainly motivated by practical reasons: A thin, straight strip of a material with sufficient elasticity can be wrapped on a surface and interpreted as a tangential strip. As a consequence, the centerline of the wrapped strip is a geodesic curve.
    Therefore, the planar grid can be fabricated from straight lamellae, and the deployment mechanism can be encoded in the planar layout.
    
    Using geodesics as grid curves is also motivated by geometry and physics. Pillwein \etal \cite{Pillwein2020} show, if the mechanics of the grid strongly correlate with the geometric properties of geodesics, the shape of the deployed grid will match the shape of the initial geodesic grid closely. 
    In the following, we briefly recall the concept of the EGG. 
    
    \begin{figure}
	\centering
	\includegraphics[width=\columnwidth]{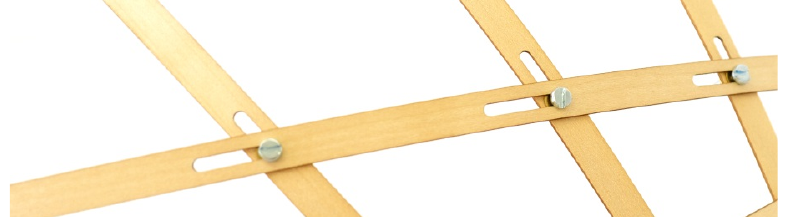}
	\caption{Notches at the connection of grid members. They enable partial sliding, which is necessary for the close approximation of the design surface. They can be implemented by elongating holes.}
	\label{fig:notch}
    \end{figure}
    
    \begin{figure*}[t]
    	\centering
    	\includegraphics[width=\textwidth]{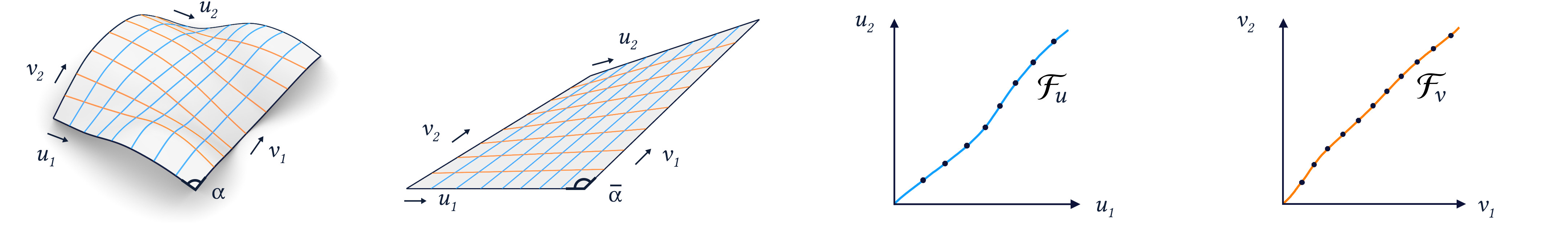}
    	\vspace{-20pt}
    	\caption{Left: A surface patch and a planar patch carrying two families of grid curves. Right: Cladding functions $\cladf_u$ and $\cladf_v$ describe all possible choices of grid members, actual grid members appear as dots in the plot. }
    	\label{fig:layout}
    \end{figure*}
    
    \subsection{Geometric Background}   \label{sec:background}
    Geodesics are curves on a surface whose geodesic curvature vanishes, i.e., their curvature is only their normal curvature on the surface. Their torsion is the geodesic torsion on the surface. 
    These properties are very beneficial for the elements of physical grids of elastic elements: 
    Bending should be easy around one axis but almost impossible around the other axis. Twisting should be easy, and stretching should be almost impossible. This behavior is inherent to thin lamellae made from a material that stretches very little. 
    The principal normals of geodesic curves coincide with the surface normals, which is also the case where two geodesics cross each other. 
    
    The goal of an EGG is to approximate a design surface in the deployed state, hence, the shape of this surface needs to be encoded in the grid layout. Therefore, a layout is computed by finding a planar grid and a geodesic grid simultaneously, which must meet the following main conditions: 
    \renewcommand{\theenumi}{\roman{enumi}}
    \begin{enumerate}
        \item Total lengths of corresponding grid curves on the surface and in the plane are equal. \label{1}
        \item Partial lengths between connections on boundary members are equal. \label{2}
    \end{enumerate}
    These conditions are based on the following idea: 
    If (\ref{1}) and (\ref{2}) are maintained and the planar grid is deployed, it is forced to take the shape of an isometry of the design surface. To match the extrinsic shape of the deployed grid to the design surface, it just needs to be bent.
    
    However, by complying with (\ref{1}) and (\ref{2}), lengths between connections of inner grid curves will not match for arbitrary surfaces. It is known in differential geometry \cite{Lagally} that in general, a spatial geodesic grid cannot be planarized by only changing the angles between grid curves. 
    The geometric error, caused by having non-matching lengths between curve-intersections, would impair the quality of the approximation. 
    To avoid this problem, the concept of sliding notches at intersections of the curves of the grid was introduced \cite{Pillwein2020}.
    They allow for a certain amount of sliding and can be implemented physically by elongated holes. 
    
    The EGG approach intertwines geometry and physics to receive curved grids that approximate design surfaces. The shape of these grids is partly driven by geometric constraints and also by the stiffness of the lamellae w.r.t. bending, twisting, and stretching.

    \begin{figure*}[t]
    	\centering
    	\includegraphics[width=\textwidth]{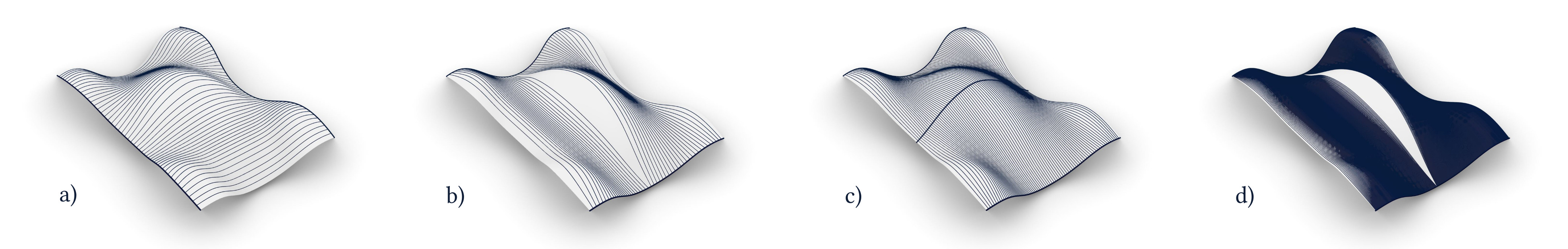}
    	\vspace{-20pt}
    	\caption{Shortest geodesics on a design surface. a) 
    	The first family of shortest geodesics covers the surface nicely and does not require splitting. b) The second family of shortest geodesics does not cover the surface completely. c) Splitting the surface into two patches enables an even coverage of the design surface. d) The problem of uncoverable regions on the surface does not depend on the number of shortest geodesics. }
    	\label{fig:unique}
    \end{figure*}
    
    \subsection{Computation of  EGG Layouts} \label{sec:approach}
    The computational aspects of designing a single elastic geodesic grid will be briefly summarized in this section; please refer to Pillwein \etal \cite{Pillwein2020} for a detailed description.
    
    The design of an EGG begins by defining the patch on the surface that will be approximated by the grid. To frame this patch, a geodesic quadrilateral is formed by picking four geodesics as boundaries.
    \stepR{The type of geodesics used to represent boundaries and grid members are shortest geodesics, i.e., the shortest paths between points on the surface.}
    
    A planar grid and the grid on the surface (cf. Figure \ref{fig:layout}) are computed simultaneously. Hence, the next step is to initialize the planar patch. It is defined by a quadrilateral in the plane with straight edges and the same side lengths as its counterpart on the surface. The planar quad has one degree of freedom, as one angle can be chosen freely to fix its shape. 
    
    Let us recall from Section \ref{sec:background}, that a grid layout needs to comply with conditions (\ref{1}) and (\ref{2}). 
    \step{Condition (\ref{2}) requires, that the parametrization of boundary members for the planar and the spatial configuration is the same. All corresponding boundary members are therefore parametrized with the same constant speed.  Figure \ref{fig:layout} shows the coordinates $u_1$ and $u_2$, which belong to the blue family of members. 
    By choosing start and end coordinates $(u_1,u_2)$, a blue geodesic on the surface and in the plane is defined. Feasible choices also need to obey condition (\ref{1}), which demands equal lengths of the geodesics on the surface and in the plane.}
    
    Finding a single grid curve maintaining (\ref{1}) and (\ref{2}) is quite simple: If we look at the setup in Figure \ref{fig:layout}, starting with an arbitrary $\overline{\alpha}$ and a combination of $(u_1,u_2)$ coordinates, $\overline{\alpha}$, $u_1$ and $u_2$ can be varied until the lengths of the geodesics on the surface and in the plane match.
    
    However, to find a global solution and cover both patches with valid grid curves, Pillwein \etal \cite{Pillwein2020} follow a different, more systematic approach: 
    \step{First, all candidates of geodesics between two opposite boundaries are described by so-called distance maps, which are functions that assign the geodesic distances to every combination of coordinates on opposite boundaries.
    Second, feasible geodesics are found by intersecting the distance maps. This process yields so-called cladding functions that are used to parametrize the corresponding surface and planar patches with geodesics (cf. Figure \ref{fig:layout}).
    
    For the blue curve-family, two distance maps, one for the planar and one for the spatial configuration, need to be computed. 
    The distance map for the surface patch $\mathcal{D}(u_1,u_2)$ is a function of the boundary coordinates $u_1$ and $u_2$, while the planar distance map $\overline{\mathcal{D}}(u_1,u_2,\overline{\alpha})$ is additionally also a function of the angle $\overline{\alpha}$. 
    
    If two distance maps $\mathcal{D}$ and $\overline{\mathcal{D}}$ for some specific $\overline{\alpha}$ are intersected, common points describe feasible corresponding geodesics on the surface and in the plane.}
    There is one degree of freedom to find good-quality grids: the angle $\overline{\alpha}$.
    Good choices of this angle deliver compact planar layouts and allow a smooth deployment of the grid. However, poor choices result in extra crossings of grid members or introduce triangles to the grid. This would destroy the kinematic mechanism, making deployment impossible.
    \step{In an optimization procedure a feasible angle $\overline{\alpha}$ is computed.
    
    Valid cladding functions $\cladf_u$ (blue curve-family) and $\cladf_v$ (orange curve-family) are bijective, establish a relationship between the boundary coordinates, and can be seen as geodesic parametrizations. 
    In other words, cladding functions enable an easy generation of a grid layout, govern the direction of geodesics and the overall appearance of the grid.} The layout can then be defined by just picking points on the cladding functions, or by using more elaborate approaches. Please refer to Pillwein \etal \cite{Pillwein2020} for more details.

    \section{Splitting the Design Surface} 
    \label{sec:modularity}
    
    \stepR{
    The objective of decomposing the design surface into multiple patches is twofold: First, if the boundary of the design surface is non-geodesic or even non-convex, using multiple patches allows for better coverage of the design surface, as a single patch may lead to big cut-offs (cf. Section \ref{sec:approach}).
    Second, if the design surface has high-$K$ regions and thus hosts non-unique shortest geodesics, it is not representable by a single EGG. A potential solution to this problem is to smooth the surface, however, we want to omit changes of the surface characteristics. The occurrence of such geodesics is strongly connected to the size of the area on the surface that should be covered by an EGG. This suggests that using multiple smaller patches is another suitable strategy to dispose of non-unique shortest geodesics. We will discuss this problem in detail and present a link between the Gaussian curvature of the surface and the uniqueness of shortest geodesics.
    
    After reviewing some aspects of differential geometry in Section  \ref{sec:geodesics_unique}, we will discuss suitable shapes for surface patches and cladding functions in Section \ref{sec:shape_planar}. In Section \ref{sec:splitting_strategy} we present an empiric method for decomposing the surface and in Section \ref{sec:distribution} we show, how to achieve $C^0$-connections of members at adjacent boundaries.
    }
    
    \subsection{Uniqueness of Shortest Geodesics}\label{sec:geodesics_unique}
    Regions of high Gaussian curvature $K$ are prone to host non-unique shortest geodesics, as Figure \ref{fig:unique} illustrates. It shows that the central region cannot be covered by shortest geodesics between the boundaries that are further apart. Moreover, one combination of points even yields two valid shortest geodesics.
    The existence of such regions and non-unique shortest geodesics is inherently connected. A check for uniqueness is known in differential geometry \cite{Carmo1992}: 
    \begin{equation} 
    \label{eq:injectivity} 
    \injR(p) \ge \frac{\pi}{\sqrt{K_\text{max}}} \,.
    \end{equation}
    The left side of the inequality is the injectivity radius $\injR(p)$ for geodesics for each surface point $p$, and $K_\text{max}$ is the maximum of the Gaussian curvature of the respective patch.
    In essence, if all geodesics on the surface patch are shorter than the value on the right-hand side of Expression (\ref{eq:injectivity}), non-unique shortest geodesics do not appear.

    Reducing the size of surface patches around a maximum of $K$ eliminates problems with non-uniqueness. \step{If the size of the patch is small enough, geodesics will be shorter than $\injR(p)$, and therefore unique. However, Expression (\ref{eq:injectivity}) is only useful as a quick check to approximately limit the size of the patch.
    }

    Clever splitting of the surface patch can reduce the problem further: Expression (\ref{eq:injectivity}) does not address the location of the peak on the patch (far away from or near a boundary). The most effective way to avoid non-unique shortest geodesics is to place the boundaries of the patches directly over the peaks. This is illustrated with a simple example, shown in Figure \ref{fig:unique} b):  
    No shortest geodesic leads over the peak, and therefore the part of the surface that is not covered can not be encoded in the grid. If the surface is split and an additional geodesic boundary leads over the peak, shortest geodesics become unique. 
    
    \begin{figure}[b]
        \begin{minipage}[c]{0.52\columnwidth}
            \includegraphics[width=\textwidth]{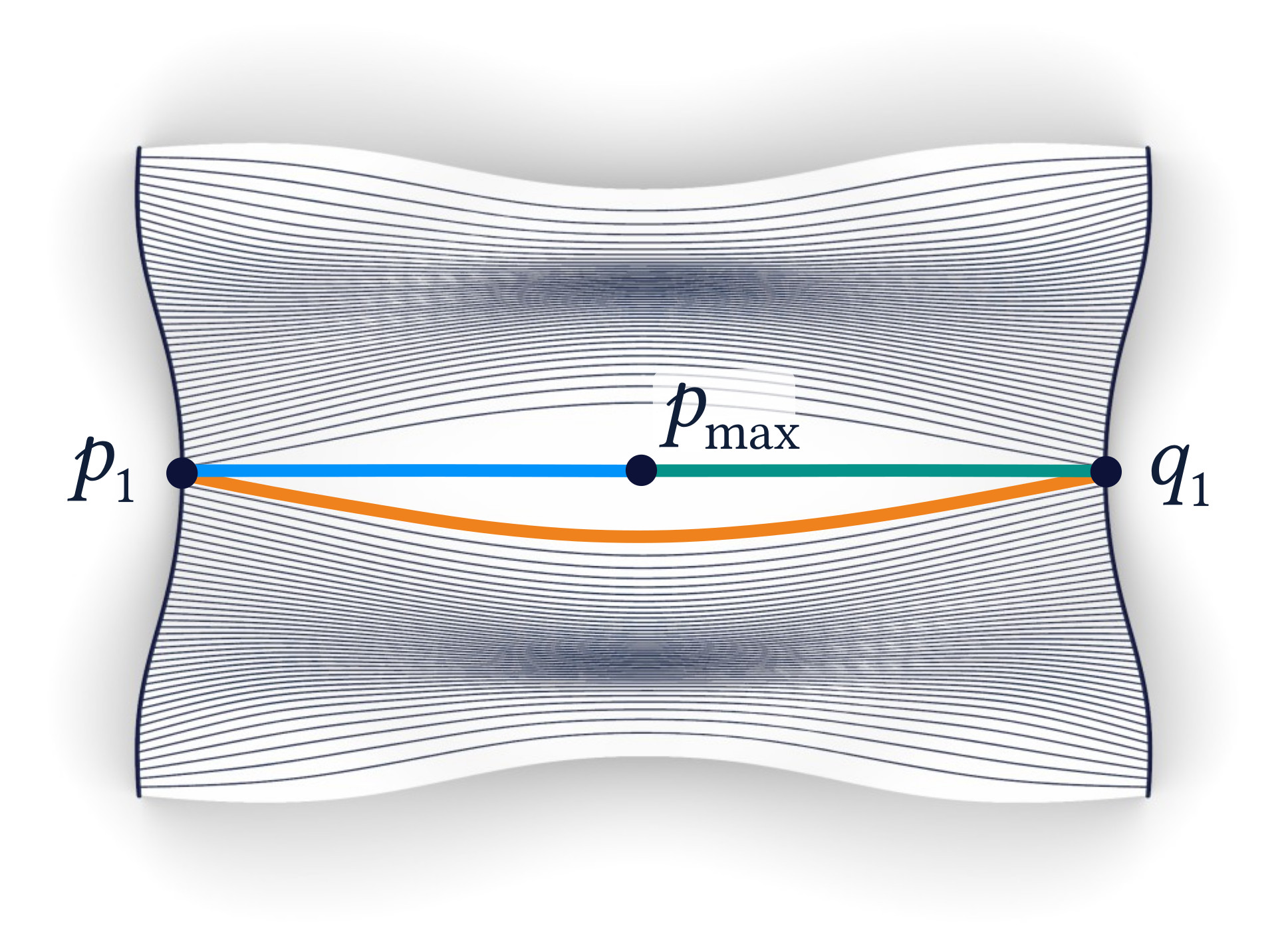}
        \end{minipage}\hfill
        \begin{minipage}[c]{0.47\columnwidth}
            \vspace{0.38cm}
            \caption{The curvature of the central peak is too high, it causes the orange shortest geodesic to omit it. We check the size of the gap indirectly by the ratio $\eta$, which is the sum of the lengths of the blue and the green geodesic, divided by the length of the orange geodesic.}
            \label{fig:eta}
        \end{minipage}
    \end{figure}
    
    \stepR{To precisely check for uncoverable areas around a Gaussian curvature peak $p_\text{max}$, we use a criterion based on geodesic distances, introduced by Pillwein \etal \cite{Pillwein2020}, which we summarize briefly. 
    Figure \ref{fig:eta} depicts the process: Imagine a geodesic through $p_\text{max}$, starting at $p_1$ and ending at $q_1$. If it is longer than the shortest geodesic connecting $p_1$ and $q_1$, this results in an uncoverable gap.

    To perform the check, we first compute geodesic distances from $p_\text{max}$ to both boundaries and one from the point $p_1$ to the opposite boundary, where $p_1$ is the closest point to $p_\text{max}$ on the first boundary.
    This provides the distances $d(p_1,q)$ to the points $q$ of the opposite boundary as well as $d(p_1,p_\text{max})$ and $d(p_\text{max},q)$. We find $q_1$ at the location of the  minimum of $d(p_1,p_\text{max}) + d(p_\text{max},q) - d(p_1,q)$. 
    If this minimum is close to zero, the peak can be covered completely. We compute the ratio of distances:
    \begin{displaymath}
    \eta = \frac{d(p_1,p_\text{max}) + d(p_\text{max},q_1)}{d(p_1,q_1)} \, ,
    \end{displaymath}
    to indirectly measure the size of the gap. Pillwein \etal propose to restrict values to a maximum of $\eta = 1.0015$.
    We use this check as an instrument to systematically decrease the size of a surface patch until uncoverable areas vanish.
    }
    
    \subsection{Shapes of Patches and Cladding Functions}\label{sec:shape_planar}
    Some combinations of shapes of the surface patch and the planar patch are undesirable in terms of the aesthetics and the functionality of the grids. Please recall, that the deployment of an EGG strongly depends on changing angles between the members (cf. Figure \ref{fig:pla_dep}). A big difference between these angles $\tod{\alpha}$ and $\alpha$ is important for smooth deployment and can be achieved if the planar patch can be "collapsed", i.e., has one very long and one very short diagonal.

    If the quad framing the surface patch has opposite sides of equal lengths, the planar quad can be collapsed perfectly to a straight line. Otherwise, there is less freedom to collapse the planar quad, as convexity must be maintained. To put it simply: The better a planar quad can be collapsed, the easier it is to fit long strips on the surface into the planar configuration.
    
    The suitability of a surface patch for an EGG can be checked by a simple geometric criterion \cite{Pillwein2020}, which supports the above considerations: 
    \begin{equation}
        \big(e-\overline{e}\big) \big(f-\overline{f}\big) < 0~,
        \label{eq:diagonals}
    \end{equation}
    where $e$, $f$ are the lengths of the geodesic diagonals on the surface patch, and $\overline{e}$, $\overline{f}$ are the lengths of the diagonals in the planar patch. If one diagonal of the planar patch is shorter than the corresponding diagonal of the surface patch, the other diagonal must be longer.
    
    \stepR{Figure \ref{fig:cladding_functions} shows how the shape of the planar patch influences the shapes of the cladding functions $\cladf$. They play a crucial role because they define the geodesic connections. Figure \ref{fig:cladding_functions} also shows, the more collapsed the planar patch is, the more "parallel" the appearance of the grid members becomes. A strong concentration of members is unfavorable for deployment, but can easily be recognized in the cladding functions, as they are linked to very steep or flat slopes. To prevent such concentrations, we set constraints on the slopes of the cladding functions as proposed by Pillwein \etal \cite{Pillwein2020} in the form of lower and upper bounds $k_\text{min}$ and $k_\text{max}$.}
    
    \subsection{Splitting Strategy}\label{sec:splitting_strategy}
    
    \stepR{Finding the best split locations is a difficult problem; it is a mix of many subjective criteria related to aesthetics (number of patches, minimum patch size, patch aspect ratio, location of supports, etc.) and geometric restrictions (unique geodesics, suitable cladding functions, etc.). 
    We provide two workflows to design a multi-patch EGG: 
    Workflow 1, presented in Algorithm~\ref{alg:workflow1}, is fully automatic, it lets a designer specify patches on the surface freely and iteratively subdivides them until all patches are feasible.
    Workflow 2, presented in Algorithm~\ref{alg:workflow2}, is a semi-automatic, more informed process and aims at keeping the number of patches low. Please note, that this process still offers a lot of freedom for the intents of a designer.
    
    \begin{figure}[t]
    	\centering
    	\includegraphics[width=\columnwidth]{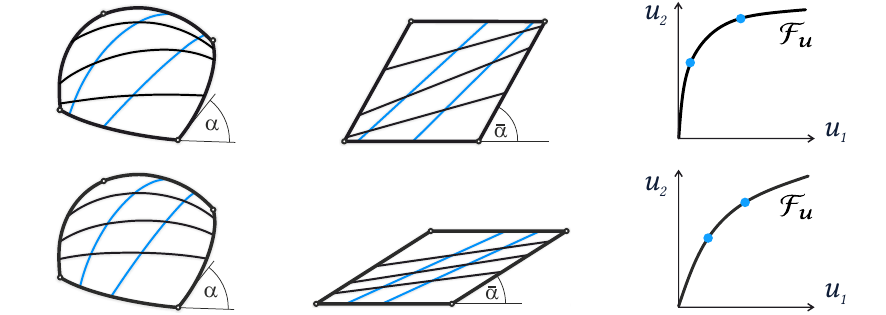}
    	\caption{ \stepR{The influence of $\overline{\alpha}$ on the shape of the planar grid and the cladding functions $\cladf$. Very steep or flat slopes of $\cladf$ indicate a concentration of members which is undesirable for the aesthetics and the deployment of the grid.} }
    	\label{fig:cladding_functions}
    \end{figure}
    
    \begin{algorithm}[!t]
        \SetAlgoLined
        Provide an initial layout of patches\;
        \While{Patches cannot be covered, $\eta > 1.0015$}{
            \uIf{$\eta>1.0015$ for one family of members}{
            Split the patch by introducing an edge over $p_\text{max}$\;
            }
            \ElseIf{$\eta>1.0015$ for both families of members}{
            Split the patch by introducing a corner at $p_\text{max}$\;
            }
         } 
        \While{Slopes of $\cladf_u$, $\cladf_v$ violate $k_\text{min}, k_\text{max}$ }{
            \uIf{$k_\text{min}, k_\text{max}$ is violated for one family}{
            Split the patch by introducing an edge over $p_\text{max}$\;
            }
            \ElseIf{$k_\text{min}, k_\text{max}$ is violated for both families}{
            Split the patch by introducing a corner at $p_\text{max}$\;
            }
         }
         \caption{Patch the Surface using Workflow 1}
         \label{alg:workflow1}
    \end{algorithm}
    
    \begin{algorithm}[!t]
    \SetAlgoLined
         \While{Not all regions of high $K$ are patched}{
            Identify remaining region of highest $K$\;
            Find $p_\text{max}$ and compute $r = \nicefrac{\pi}{\sqrt{K_\text{max}}}$\;
            Draw a circle ($p_\text{max}$, $r$) on the surface\;
            \eIf{The size of the circle is suitable for a patch}{
            Define a patch with corners on the circle\;
            }{
            Introduce a corner at $p_\text{max}$, use the directions of the principal curvatures $\kappa_1$ and $\kappa_2$ as guides for the directions of the patch edges\;
            Draw a patch which is split by the edges through $p_\text{max}$ into four parts, each of them corresponding to the size of the circle\;
            Grow the patches outwards iteratively until either the patches are big enough for designer demands or $\eta \ge 1.0015$\;
            Check if two adjacent patches can be merged using $\eta$\;
            }
            \If{Slopes of $\cladf_u$, $\cladf_v$ violate $k_\text{min}, k_\text{max}$}{
            Decrease the size of the patch to make the members of the respective family shorter until $k_\text{min}, k_\text{max}$ are met\;
            }
        } 
         
        \While{Surface is not fully patched}{
            Choose a region in a greedy manner\;
            Use similar sized patches in regions with higher $K$\;
            Use larger patches in regions with low $K$\;
            \If{Slopes of $\cladf_u$, $\cladf_v$ violate $k_\text{min}, k_\text{max}$}{
            Decrease the size of the patch to make the members of the respective family shorter until $k_\text{min}, k_\text{max}$ are met\;
            }
        } 
         \caption{Patch the Surface using Workflow 2}
         \label{alg:workflow2}
    \end{algorithm}

    Let us first summarize problems that induce a split: 
    The cut-offs of the design surface using a single EGG are too big. There are non-unique shortest geodesics, i.e., some areas of the surface patch cannot be covered by shortest geodesics (cf. Section \ref{sec:geodesics_unique}).
    Grid members are too concentrated in certain areas of the surface patch, i.e., the slopes of the cladding functions are too steep or too flat (cf. Section \ref{sec:shape_planar}).

    Both workflows have the objectives of ensuring full coverage of the patches and suitable cladding functions. We check them using $\eta$, $k_\text{min}, k_\text{max}$, and adjust the patches by splitting them or decreasing their size until they are feasible. Workflow 2 is based on defining the patches in regions of the highest $K$ first because they are prone to uncoverable areas and have a high aesthetic impact on the overall design. After that, the remaining parts of the surface are patched in a greedy manner.
    }

    \subsection{Distribution of Members in a Multi-Patch EGG}\label{sec:distribution}
    
    \begin{figure}[t]
    	\centering
    	\includegraphics[width=0.9\columnwidth,trim=0 18 0 0, clip]{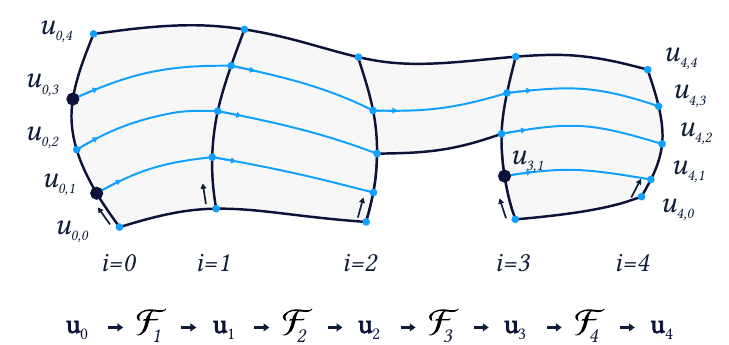} 
    	\caption{ \stepR{ Members of the blue family are propagated using the cladding functions $\cladf_1, \cladf_2, \cladf_3, \cladf_4$. Black dots represent $u$-values that define members, blue dots indicate how members are propagated. If there are smaller patches within a family of grid members, like patch number 3, they define an original member. Please note, that back-propagating members is performed using $\cladf_i^{-1}$.}
    	}
    	\label{fig:members}
    \end{figure}

\begin{figure}
        \centering
        \begin{subfigure}{0.49\columnwidth}
            \centering
            \includegraphics[trim=6.5 32 25 54,clip,width=0.99\textwidth]{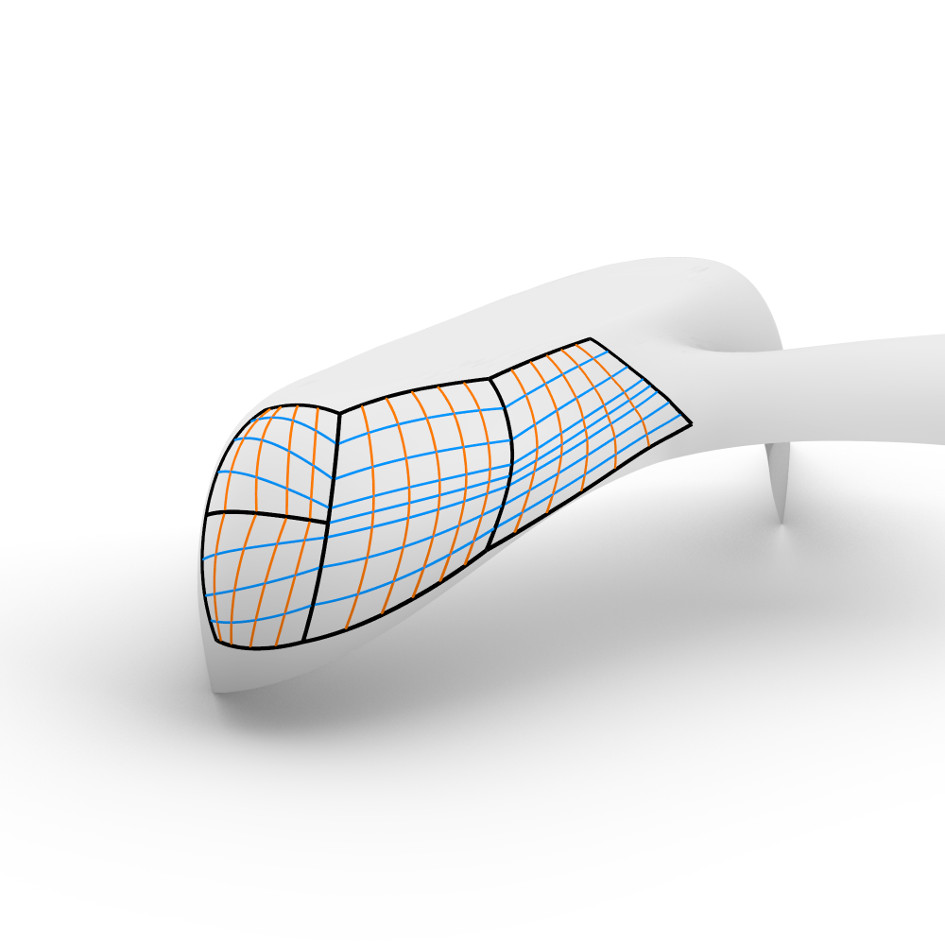}
        \end{subfigure}
        \hfill 
        \begin{subfigure}{0.49\columnwidth}
            \centering
            \includegraphics[trim=6.5 32 25 54,clip,width=0.99\textwidth]{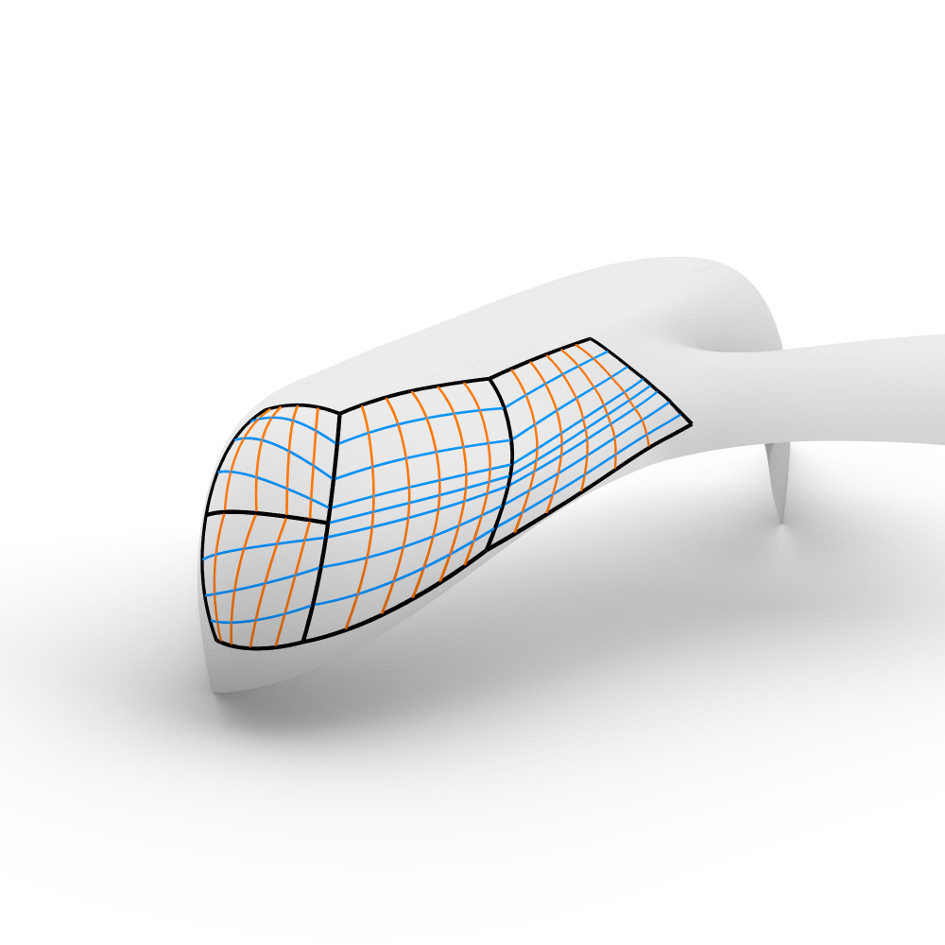}
        \end{subfigure}
        \caption{Distribution of members on a multi-patch EGG design. Left: Members are distributed evenly on the leftmost boundary and simply propagated to the other patches. Right: The evenness of members is improved by solving the Optimization Problem (\ref{eq:members}). }
        \label{fig:distribution}
    \end{figure}

    \stepR{
    To create an aesthetically pleasing multi-patch EGG with best coverage,  
    we distribute members equidistantly across each patch and make sure that connections at adjacent patch boundaries maintain $C^0$-continuity.  
    
    A member of the $u$-family is defined by its start and end coordinates $u_1$ and  $u_2 = \cladf_1(u_1,\tod{\alpha})$ respectively. 
    The subsequent members in following adjacent patches can be denoted at their boundaries as $u_3 = \cladf_2( u_2, \tod{\alpha}_2)$, etc. 
    Using this notation, we formulate an optimization problem, where we minimize the normalized squared distances between member endpoints on the boundaries: \begin{align}
        \min_{\mathbf{u}_i} \sum_{i=1}^n  \omega_i \, \langle \Delta \mathbf{u}_i ,\, \Delta \mathbf{u}_i \rangle \,,
        \label{eq:members}
    \end{align}
    \text{with} 
    \begin{equation*} 
         \Delta \mathbf{u}_i = \frac{1}{l_i} \bmat{ u_{i,1} -u_{i,0} \\ u_{i,2} -u_{i,1}\\ ...\\  u_{i,m} -u_{i,m-1}}\,,
    \end{equation*}
    where $n$ is the number of boundaries, $i$ is the index of the current boundary, $m-1$ is the number of members, $l_i$ is the length of the respective boundary, $\omega_i$ is an importance factor for the respective boundary, and $u_j, \dots , u_{m}$ are the coordinates of start or endpoints on the respective boundary. Please note, that the second family of members is expressed using $v$ in our notation (cf. Figure \ref{fig:layout}) and solved independently. 
    
    We solve Problem (\ref{eq:members}) using a gradient-based sequential quadratic programming algorithm in \textsc{Matlab}. The number of members is fixed during the optimization, members due to boundaries of smaller patches appear as constraints as depicted in Figure \ref{fig:members}. 
    It also shows a family of members on multiple patches, the variables for Problem (\ref{eq:members}) are the $u$-values, displayed as black dots. Other $u$-values are computed by propagating members using the cladding functions.
    Figure \ref{fig:distribution} shows the result of the optimization on a set of patches.}

    \section{Fabrication}\label{sec:challenges}
    There are challenges associated with the design and the implementation of EGG on different scales. This section presents some of the challenges and possible solutions. The showcase model shown in Figures \ref{fig:side} and \ref{fig:pavilion} was used to evaluate these approaches. 
    The measures presented are intended to improve the feasibility of EGG on an applied level.
    
    \subsection{Notches and Friction} \label{sec:friction}

    \begin{table}[b]
     \caption{Empirically determined friction coefficients $\mu$ for static and sliding friction; the evaluated plywood is poplar.}
        \begin{tabular}{cccc}
           Material &    ~~~~~~~~~$\mu_{static}$ ~~~~~~~~~~~ &  ~~~~ $\mu_{sliding}$~~~~~~~~\\[0.02cm]
        \midrule
       Plywood-Plywood   & 0.436                         & 0.273                  \\
        PTFE-PTFE       & 0.163                         & 0.091                   \\
        Plywood-Steel    & 0.252                         & 0.203                   \\[0.02cm]
        
        \end{tabular}
        
        \label{tab:mu}
    \end{table}

    The quality of the approximation strongly depends on the concept of notches, as outlined in Section \ref{sec:background}. Notches enable grid members to slide for a predefined length and direction at their connections. Figure \ref{fig:notch} shows notches in a physical grid.

    While sliding of members in the notches delivers perfect results in simulations, friction poses a problem for the sliding process of physical grids. lamellae often get stuck in the notches, because forces that are supposed to push the lamellae in the right direction inside the notches are counteracted by friction.
    By interfering with sliding, the quality of the approximation suffers, and local stresses arise as some lamellae buckle. Also, the dissipation of energy by friction increases the force necessary to deploy the grid.
    
    The regions that contribute to the overall friction during deployment are the contact areas of lamellae, the screw threads, and the notches, and the lamellae and the washers. Table \ref{tab:mu} summarizes empirically determined friction coefficients. The goals of easy deployment on one hand and secure final fixing, on the other hand, can be fulfilled by reducing the friction between the lamellae but maintaining friction between lamellae and washers.
    
\begin{figure}[t]
    	\begin{minipage}[b]{0.235\textwidth}
    		\centering
    		\includegraphics[width=\textwidth]{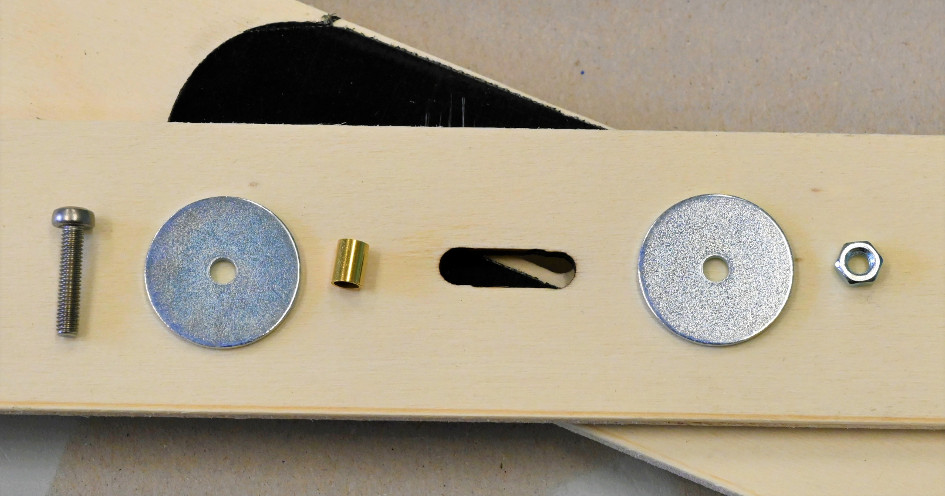}
    		\end{minipage}
    		\hfill
    	\begin{minipage}[b]{0.235\textwidth}
    		\centering
    		\includegraphics[width=\textwidth]{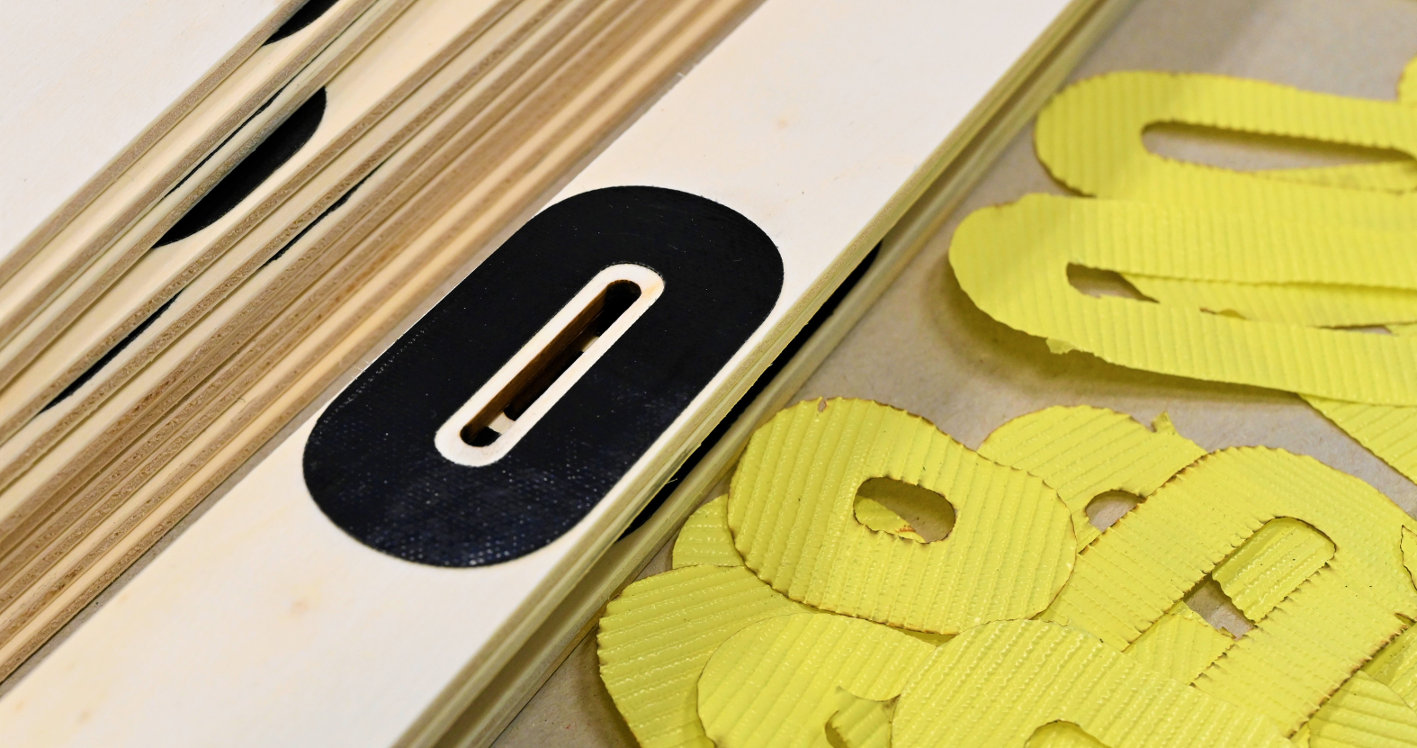}
    	\end{minipage}
    	\vspace{2 pt}
    	\caption{Design features to minimize friction. Left: Contact of the screw thread and the notch wall is prohibited by a brass shell. Right: PTFE-foil stickers reduce friction at lamella-to-lamella contact areas.}
    	\label{fig:friction}
    \end{figure}

    An effective measure to reduce friction is to equip the lamella-to-lamella contact areas with a PTFE-foil layer. The friction coefficient of PTFE is quite low, and it is robust. For every notch, an individual PTFE sticker was cut from an adhesive foil. The stickers cover the immediate contact area for every notch, so there is no contact with the rough wooden surface of the lamellae. 
    Brass shells furthermore reduce the friction of the screw in the notch. Figure \ref{fig:friction} illustrates the measures taken to reduce friction.
    
    \subsection{Grid Size}\label{sec:scaling}
    
    Fabricating small prototypical models to investigate the qualities of the design is common practice in architecture. Once a suitable material is found, designing EGG models on the same scale is easy. 
    
    Analyzing existing models \cite{Pillwein2020}, we provide the rough empiric rule for small and medium-sized models, that a lamella-width of around two percent of the desired output size works well. 
    The thickness of a lamella follows from the considerations in Section \ref{sec:background} as smaller than $\nicefrac{1}{5}$ of the width. Please recall, that a distinct width-to-thickness ratio is necessary for the grid to behave like a geodesic grid.
    Building EGG models on a larger scale is not straightforward, which will be discussed in this section. 
    
    The shape of a deployed elastic grid is driven by the stiffness parameters of the lamellae. The related X-Shells approach \cite{Panetta2019} even uses different shapes of cross-sections, which induce different stiffness parameters to influence the shape of a deployed X-Shell. 
    The stiffness of an element like a lamella can be expressed w.r.t. the different types of deformation, like bending and twisting. It determines the amount of stress which is induced by the deformation. The magnitude of a stiffness parameter is set by the geometry of the cross-section and by the material parameters. 
    In our case, when wrapping a lamella on the surface patch, bending and twisting are prescribed. 
    
    To get an insight into the matter, we will first consider scaling up the whole structure just linearly, including the cross-sections of the lamellae, and have a look only at the internal stresses caused by bending.
    
    Let us further assume, that the lamellae of the deployed grid agree well with the scaled surface patch, no matter how high the scaling factor is. This enables us to compute the stresses due to bending w.r.t. a scaling factor $f$:
    \begin{align*} 
    \sigma_{B,max}(f) = &  \pm \frac{M(f)}{I(f)} \frac{tf}{2}= \pm \frac{E~ I(f) \frac{\kappa}{f} }{I(f)} \frac{tf}{2} =   \pm \frac{E \kappa  t}{2} \,,
    \end{align*}
    where $\sigma_{B,max}$ is the maximum stress induced by bending, $M$ is the bending-moment, $I$ is the moment of inertia, $\kappa$ is the curvature, and $t$ is the thickness of the lamellae.
    This means that the bending stresses are constant under linear scaling, as the curvature decreases at the same rate as the thickness of the cross-section increases.
    
    \begin{figure}[t]
    	\centering
    	\includegraphics[width=\columnwidth]{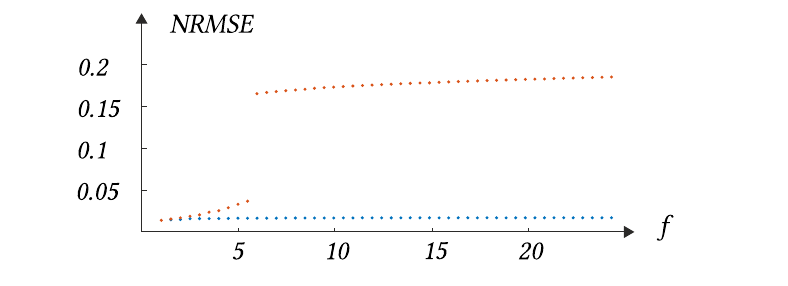}
    	\vspace{-12pt}
    	\caption{The effects of scaling an EGG. The NRMS-Error is used to measure the deviation of scaled simulated grids to the design surface.
    	Blue dots represent simulation results, if gravity is neglected, orange dots represent results including gravity. The jump represents the collapse of the structure under its own weight.}
    	\label{fig:scaling_f}
    \end{figure}
    
\begin{figure}[t]
    	\centering
    	\includegraphics[width=0.93\columnwidth]{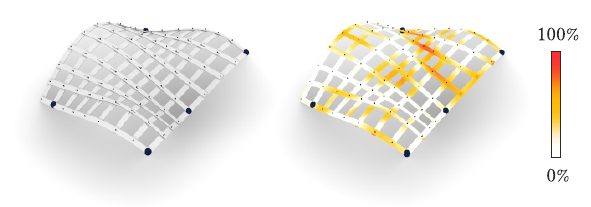}
    	\caption{Simulated elastic grid and deviations to the design surface. The deviations relate to the Euclidean distance w.r.t. the width of a lamella.
    	The mean deviation is 2 mm for absolute dimensions of 0.43 x 0.57 x 0.1 m.}
    	\label{fig:deviations}
    \end{figure}

    The shape of a model is determined by the geometric boundary conditions (deploying a grid and fixing it to supports) and stiffness. This interplay leads to internal stresses. For EGG they are mainly caused by bending the lamellae and their self-weight. As the scaling factor $f$ increases, stresses caused by self-weight grow significantly, whereas stresses caused by bending remain constant.
    The strong increase of stresses caused by self-weight is due to the fact that it grows cubically w.r.t. the scaling factor $f$: 
    \begin{align*} 
    F_G(f) =  &  m(f)  g = V(f)  \rho  g =  t  w  l  f^3  \rho  g \,,
    \end{align*}
    where $w$ is the width and $l$ is the length of a lamella.
    Figure \ref{fig:scaling_f} shows numerical results of the impact of scaling on the shape, generated with a physical simulation using the Discrete Elastic Rods model \cite{Bergou2008, Bergou2010}. The implementation we used is based on the implementation by \cite{Vekhter2019}, which features the simulation of self-weight.
    To compare the shapes of linearly scaled versions of the small grid in Figure \ref{fig:pla_dep}, the Normalized Root Mean Square Error ($\NRMSE$) was used. The difference between the predicted and the observed values in the $\NRMSE$ are the distances $d_i$ between points on the centerlines of a simulated deployed grid and their nearest neighbors in the surface patch. The inverse scaling factor acts as the normalization factor: 
    \begin{align*} 
    \NRMSE = \frac{1}{f}\sqrt{ \frac{1}{n} \displaystyle \sum_{i=1}^{n} d_i^2  } ~.
    \end{align*} 
    Figure \ref{fig:scaling_f} shows, that linear scaling does not influence the shape significantly, as long as gravity is neglected. However, taking gravity into account, the cubical increase of gravitational loads makes linear scaling feasible only for small scaling factors. 
    Figure \ref{fig:deviations} shows the simulated shape of the desktop model, which corresponds to the leftmost data point in Figure \ref{fig:scaling_f} ($f=1$).

    \begin{figure}[t]
    	\centering
    	\includegraphics[width=0.91\columnwidth]{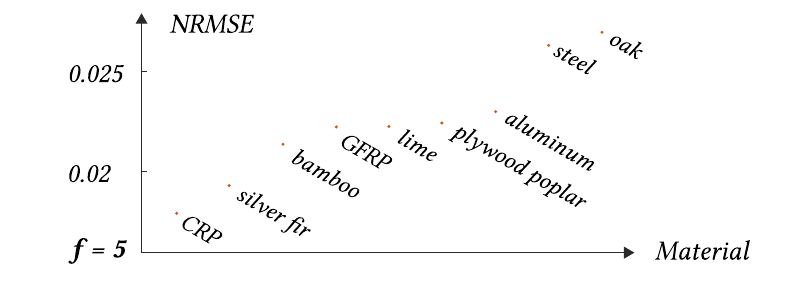}
    	\caption{The performance of different materials at $f=5$. The plot shows that fiber-reinforced polymers or timbers are quite suitable. High-density materials like steel or oak perform worse, although they have a high elastic modulus.}
    	\label{fig:materials}
    \end{figure}

    \paragraph{Strategies for large Grids}
    To enable the construction of larger structures, the gravitational forces need to be kept small, while we expect a positive effect on the shape when increasing the relevance of bending stresses.
    \step{There are several options to tackle this problem:
    \begin{enumerate}
        \item Choose suitable materials. \label{en:material}
        \item Change the grid design by introducing more members with smaller cross-sections. \label{en:members}
        \item Introduce multi-layered structures. \label{en:multi}
        \item Introduce a support at every connection on the boundary. \label{en:boundary}
    \end{enumerate}
    We will subsequently discuss approaches (\ref{en:material}) and (\ref{en:members}) in detail. Approach (\ref{en:multi}) is referred to future work and approach (\ref{en:boundary}) is straightforward.
    }

    Changing the material allows an improvement in two ways: Making it stiffer by increasing the elastic modulus $E$, and making it lighter by reducing the density $\rho$. It is therefore practical to look for a material with the right ratio of the two.
    In fact, this ratio is called specific modulus $\lambda = \nicefrac{E}{\rho}$ and is well known in light-weight engineering like the aerospace design. There, it is used for parts whose shape is driven by stiffness, like the wings. Figure \ref{fig:materials} shows the impact of using different materials on the shape of an initially desktop-sized EGG when scaling it by a factor of $f=5$.
    However, when increasing the elastic modulus, there is a restriction that needs to be considered. The internal stresses due to bending $\sigma_{B}$ grow proportionally, as the curvature $\kappa$ and the thickness $t$ of the lamella are prescribed.
    Therefore, it needs to be checked, that the stresses do not exceed the strength of the material.

    Another aspect for the choice of material, is caused by geometry: As the ribbon-like lamellae are wrapped onto the surface, they need not only to bend but to twist as well. In extreme cases, on surface regions of negative Gaussian curvature, lamellae may even be almost or completely straight, but still have to twist. 
    \stepR{If such a lamella is made from material that allows no in-plane stretching, it needs to buckle. 
    Using materials like wood does not cause trouble: the fibers are approximately perpendicular to the cross-section, and the outer fibers can stretch minimally w.r.t. the fibers close to the centerline. Recent research on grid structures made of thin lamellae that undergo large torsion \cite{schikore2019} even shows, that metal is a suitable material. 
    }
    
    Scaling a structure non-uniformly is feasible only within certain width-to-thickness bounds, as mentioned earlier. Increasing the thickness without keeping a high width-to-thickness ratio will corrupt the EGG design concept, and thus worsen the quality of the approximation.
    
    \step{Large grids, spanning meters to tens of meters, can furthermore not be obtained by simply scaling a desktop prototype grid. The layout needs to be adapted, which means introducing more and smaller lamellae. This can be explained simply by considering the effects of gravity: Doubling the number of lamellae doubles the weight of a grid. Scaling the lamellae by a factor of two increases the weight by a factor of eight.
    
    The insights on shape and material suggest, that the bigger an EGG gets, the more important high-quality material and the dimensions of the cross-sections become. Good material choices can be achieved by tuning the specific modulus $\lambda$ to receive the best possible shape. However, finding the best cross-sectional dimensions and number of lamellae for large grids is a difficult engineering task that is out of the scope of this paper.
    }

    \begin{figure*}[t]
    	\centering
    	\includegraphics[width=0.98\textwidth]{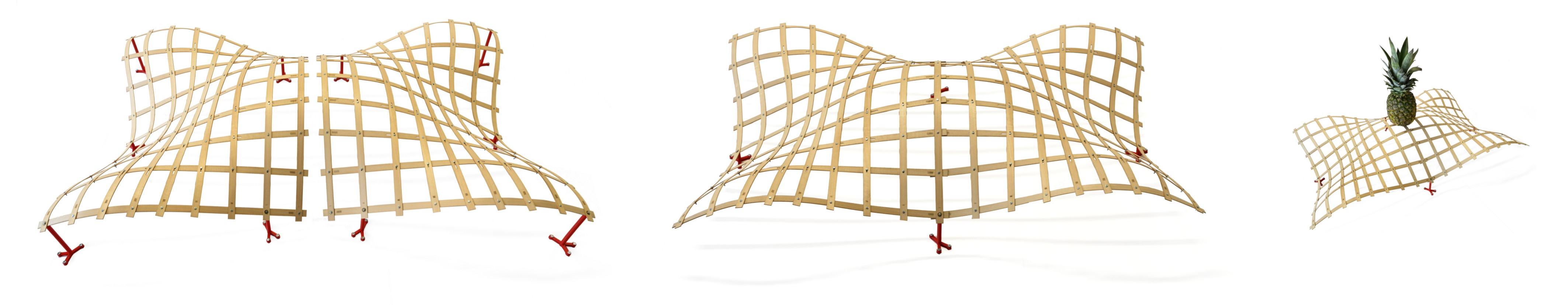}
    	\caption{Modules of the small-scale prototype and a test of the structural behavior. Left and middle: Modules can be used separately or combined. Right: A test of the load-bearing capacity of the grid. }
    	\label{fig:small_scale}
    	\label{fig:structural}
    \end{figure*}

    \subsection{Deployment} \label{sec:deploy}

    In theory, an EGG can be deployed to match all isometries of the input surface patch, which is due to the design concept. If the deployment is not guided, the grid will take the shape corresponding to minimal elastic energy. This shape will most certainly not correspond to the shape of the input surface patch.
    
    To fix this, first, the deployment process of the grid needs to be guided.
    This means that the shape induced by deployment needs to be checked and adjusted, to ensure that the grid buckles in the right way.  
    To put it simply: All bumps, that are supposed to go up, actually go up. Adjusting the shape of the grid is easy at the beginning of the deployment, as elastic forces are still low.
    Depending on the complexity of the design surface, however, this step can be tricky. So, for large and complex designs, it makes sense to set a maximum patch size w.r.t. handling of the grid. Multi-patch EGG can be deployed and combined sequentially, which simplifies this task.
    
    Second, after deploying, the grid needs to be bent to match the shape of the surface patch. To ensure this, a set of supports is used. Good choices for locations are intersections of grid curves with boundaries. A support is defined by a point and a plane: The point corresponds to the intersection point and the plane corresponds to the tangent plane at this point.

    Adjacent grids can be simply connected along their common lamellae, as can be seen in Figure \ref{fig:small_scale}. 
    Please note, that there are no notches on boundary members.

    \section{Results}\label{sec:pavilion}
    
    \step{In this section, we present models, designed with the EGG approach, using multiple patches. 
    First, the whole design and fabrication process of a  showcase model is presented.
    Second, some iconic design surfaces from free-form architecture are used to explore the potential of the multi-patch EGG approach on an architectural scale.
    }

    \subsection{Showcase Model}
    As a starting point for a model of a few meters in size, a small-scale prototype was built. It was intended to assess the aesthetic and structural qualities of the design.
    The first attempt to build the full-scale model failed, because of poor material properties. The second attempt, however, was successful. The production process was automated as much as possible, e.g. by laser-cutting the notches and lamellae.

\begin{figure}[b]
	\centering
	\includegraphics[width=\columnwidth]{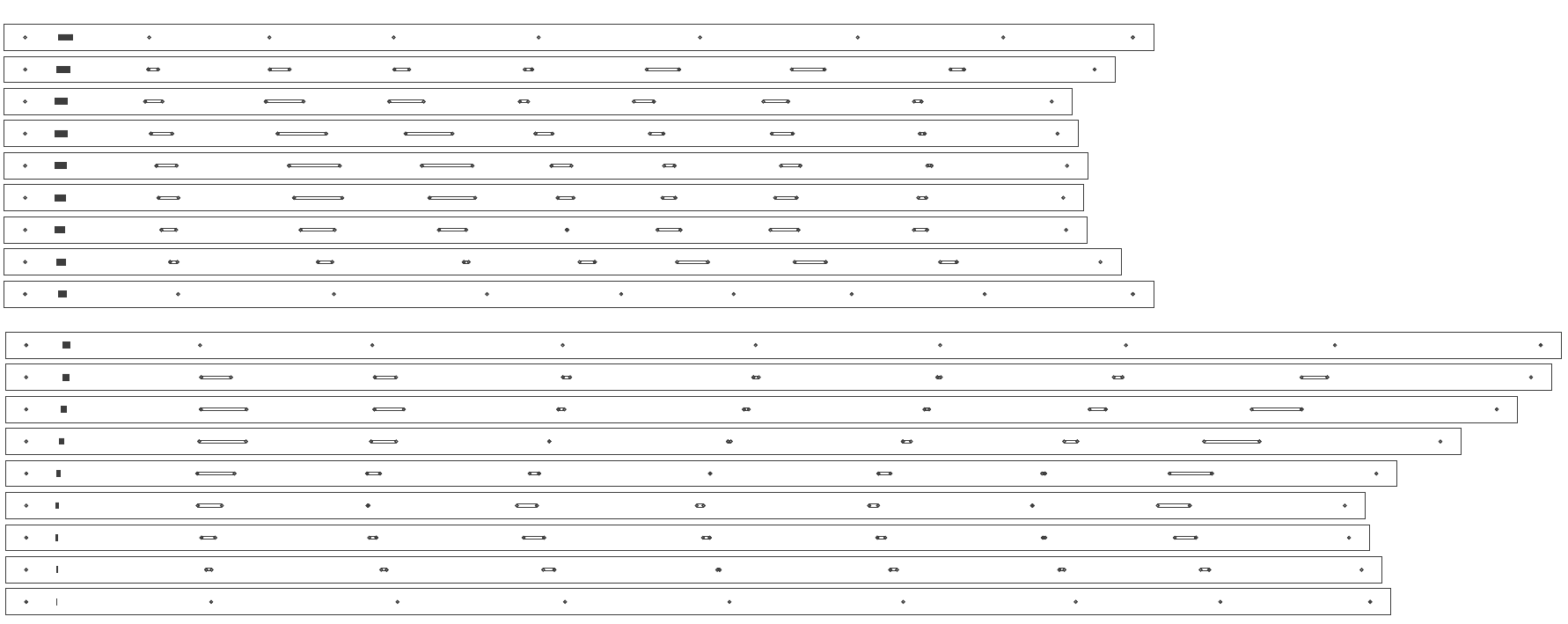}
	\caption{Plan for laser-cutting the lamellae. Scaling the plan by a factor of 9 yields the size of the prototype of Figure \ref{fig:pla_dep}.}
	\label{fig:plans}
    \end{figure}

    \paragraph{Prototype}
	The design was created by an architecture student without particular prior knowledge of elastic structures or differential geometry. The student was given access to the EGG pipeline via a Grasshopper node.
	The simulated shape of the current design and the laser-cutting plans were output again in Grasshopper.
	Out of many design candidates we chose the structure of Figure \ref{fig:structural}, which features two symmetric modules that can be used separately or together. The measurements of the prototype were 0.85 x 0.57 x 0.15 meters, and it was built from 1 mm thick and 10 mm wide limewood lamellae. A plan for laser-cutting the lamellae is provided in Figure \ref{fig:plans}. The 3D-printed supports feature inclined contact areas. The prototype kept its shape even with as few as four supports, which was a relevant design feature.
    Tests suggested a high degree of structural stiffness and load-bearing capacity.
    The total weight of the model was 160 grams, the applied weight in Figure \ref{fig:structural} was 1135 grams. This gives a promising load-to-self-weight ratio of about 7. 
    
    \begin{table}[b]
        \caption{The specific modulus $\lambda$ of the materials used ($\perp$ and $\parallel$ indicate fiber direction). Limewood was used for the small-scale prototype, the failed version was built from birch plywood, and the successful medium-scale model was built from poplar plywood.}
        \begin{tabular}{ccccc}
            &&Limewood& Plywood & Plywood   \\
            Parameter & Unit    &   $\parallel$ &birch $\perp$  & poplar  $\parallel$\\[0.05cm]
            \midrule
            $E$    &      [$GPa$]                   & 9.1   & 4.0   & 7.6            \\
            $\rho$ &[$\nicefrac{g}{cm^3}$]          & 0.50   & 0.65  & 0.43  \\
            $\lambda=\nicefrac{E}{\rho}$&[$10^{6} \nicefrac{m^2}{s^2}$]  & $18.2$ & $6.15$ &  $17.7$ 
        \end{tabular}
        
        \label{tab:specific}
    \end{table}
    
\begin{figure*}[t]
    \centering
	\includegraphics[width=\textwidth]{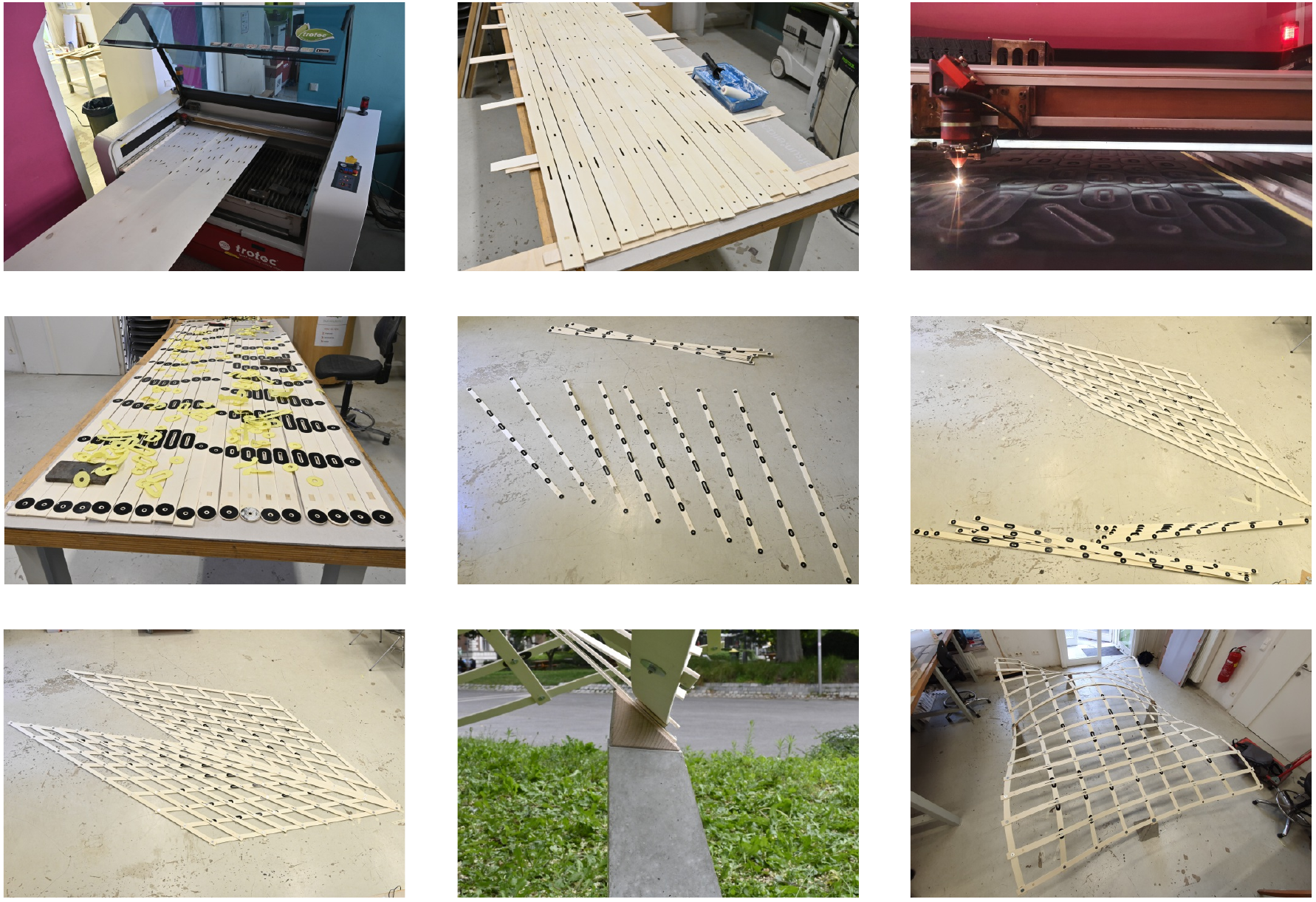}
	\caption{Steps of the fabrication process of the model. A laser-cutter was used to produce the lamellae and the PTFE-stickers. After sticking them to the lamellae, the modules of the grid were assembled. The supports were cast from concrete and have inclined contact areas. Finally, the grid was deployed and fixed to the supports.}
	\label{fig:process}
    \end{figure*}

	\paragraph{Fabrication Process}
	Fabricating the full-scale model featured the following steps: laser-cutting the lamellae and the notches, sanding and coating the lamellae, cutting the PTFE-stickers, assembling the grid, and casting the supports. The different steps of the fabrication process can be seen in Figure \ref{fig:process}.
	
	As a construction material, plywood was used. It is cheap, available in big panels, and easy to machine. The plywood we used had three layers and was 3 mm thick. We decided to use a laser-cutter to produce the lamellae and notches.
	Our Trotec Speedy 500 has a back flap that can be opened. This enables easy production, as the plywood panels could be pushed through the flap. 
	The cutting happened in multiple steps: after cutting one segment, the plywood panel was pushed forward and adjusted to cut the next segment.  
    
    The supports were cast from concrete and designed to be stable under their own weight, but also light enough that they could be carried.
    The inclined contact areas of the supports were realized with wooden wedges mounted to the supports.

	\begin{figure}[b]
    	\begin{minipage}[b]{0.68\columnwidth}
    	\centering
		\includegraphics[width=0.8\textwidth]{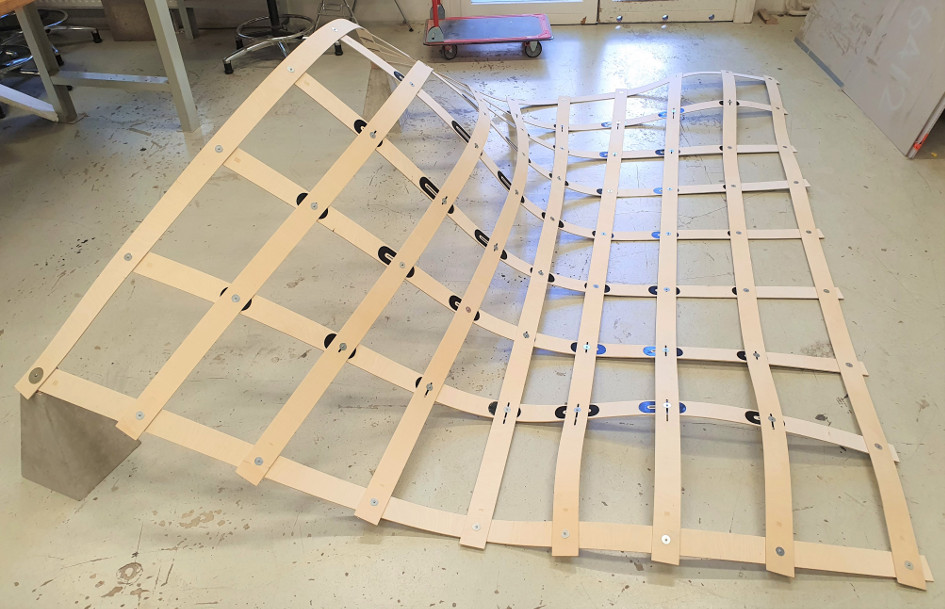}    		\end{minipage}
     	\begin{minipage}[b]{0.3\columnwidth}
    	\centering
		\includegraphics[width=0.89\textwidth]{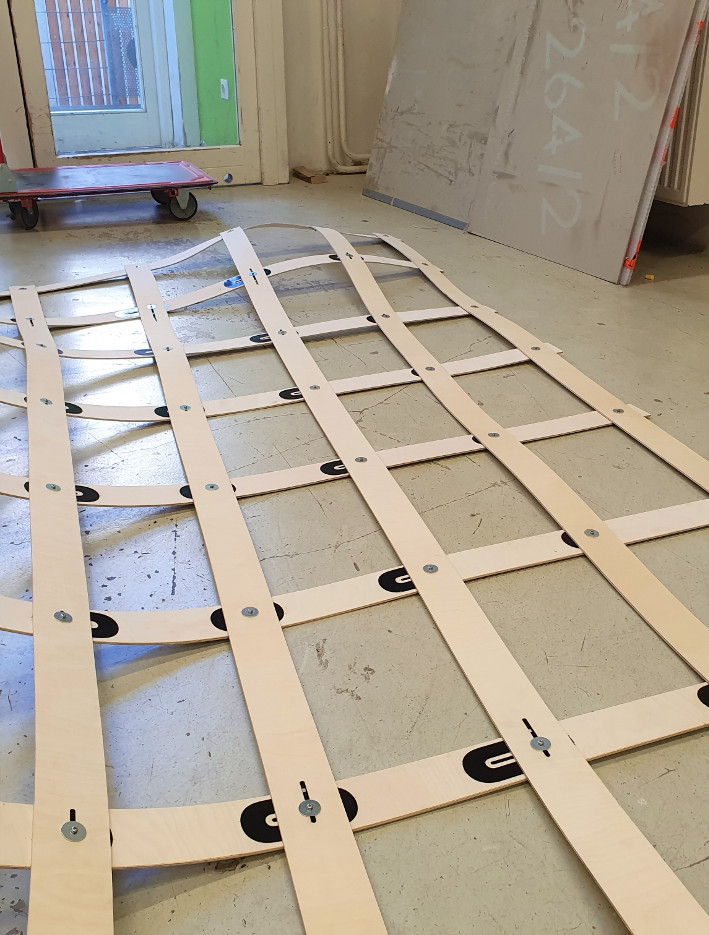}
    	\end{minipage}
    	\caption{One module of the failed medium-scale model. The structure is not able to carry its self-weight. Some curvature features can be recognized, however, the structure as a whole performs rather poorly.}
    	\label{fig:fail}
    \end{figure}
	
	\paragraph{Failed Model}
	In the first iteration of building the full-scale model, the change of material from high-quality limewood to simple plywood had negative side effects. The structure failed under its own weight due to very flexible lamellae. The reason for the flexibility was the wrong fiber orientation of the top layers of the plywood. This reduced the elastic modulus $E$ to about $\nicefrac{1}{5}$ of what was expected ($E$ of timber perpendicular to fiber direction is about $\nicefrac{1}{5}$ of $E$ parallel to fiber direction). 
	Therefore, the top layers hardly contributed to the structural performance but made up $\nicefrac{2}{3}$ of the self-weight.

	Table \ref{tab:specific} summarizes the values of the specific modulus for the materials we used.
	Comparing the specific modulus of plywood (birch) to limewood, it is only about $\nicefrac{1}{3}$.
	Essentially, the plywood was way less performative and much heavier, which is exactly the opposite of what would have been appropriate for a larger grid (cf. Section \ref{sec:scaling}).
	
\begin{figure}
	\centering
	\includegraphics[width=0.98\columnwidth]{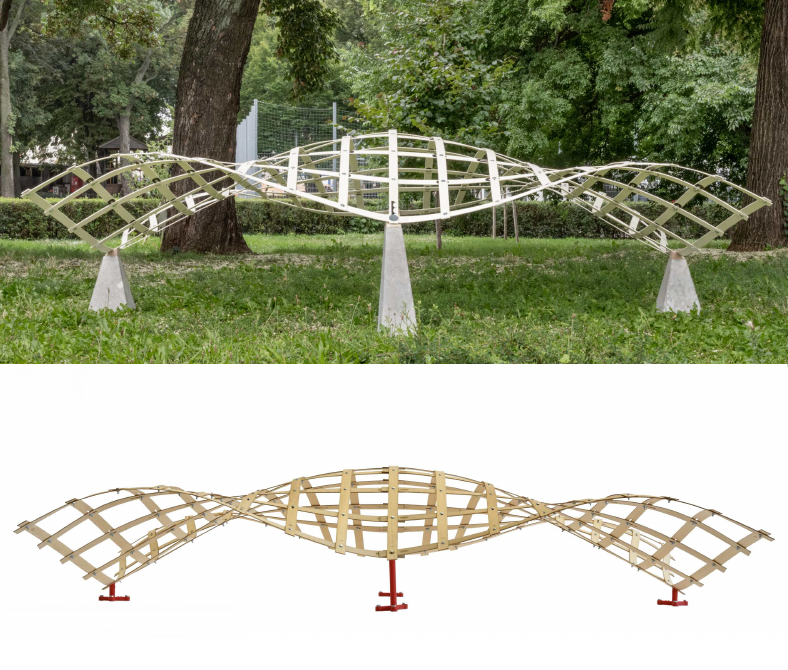}
	\vspace{-20pt}
	\caption{Side-view comparison of the shapes of the small-scale prototype and the showcase model. Good agreement of the shapes is obvious.}
	\label{fig:side}
    \end{figure}

	\paragraph{Successful Model}
	The second attempt to build the full-scale model succeeded, using poplar plywood for the lamellae. This plywood is quite efficient, as Table \ref{tab:specific} shows. 
	In the deployed state the structure measures 3.1 $\times$ 2.1 $\times$ 0.9 meters (including supports) and has a self weight of 7.1 kilograms. This makes a weight-to-span ratio of $1.09 ~ \nicefrac{kg}{m^2}$ and a thickness-to-span ratio of $\nicefrac{1}{516}$. 
	In Figure \ref{fig:side}, the closeness of the shapes of the small-scale prototype and the medium-scale model is obvious.
	
\begin{figure*}
	\centering
	\includegraphics[width=0.99\textwidth]{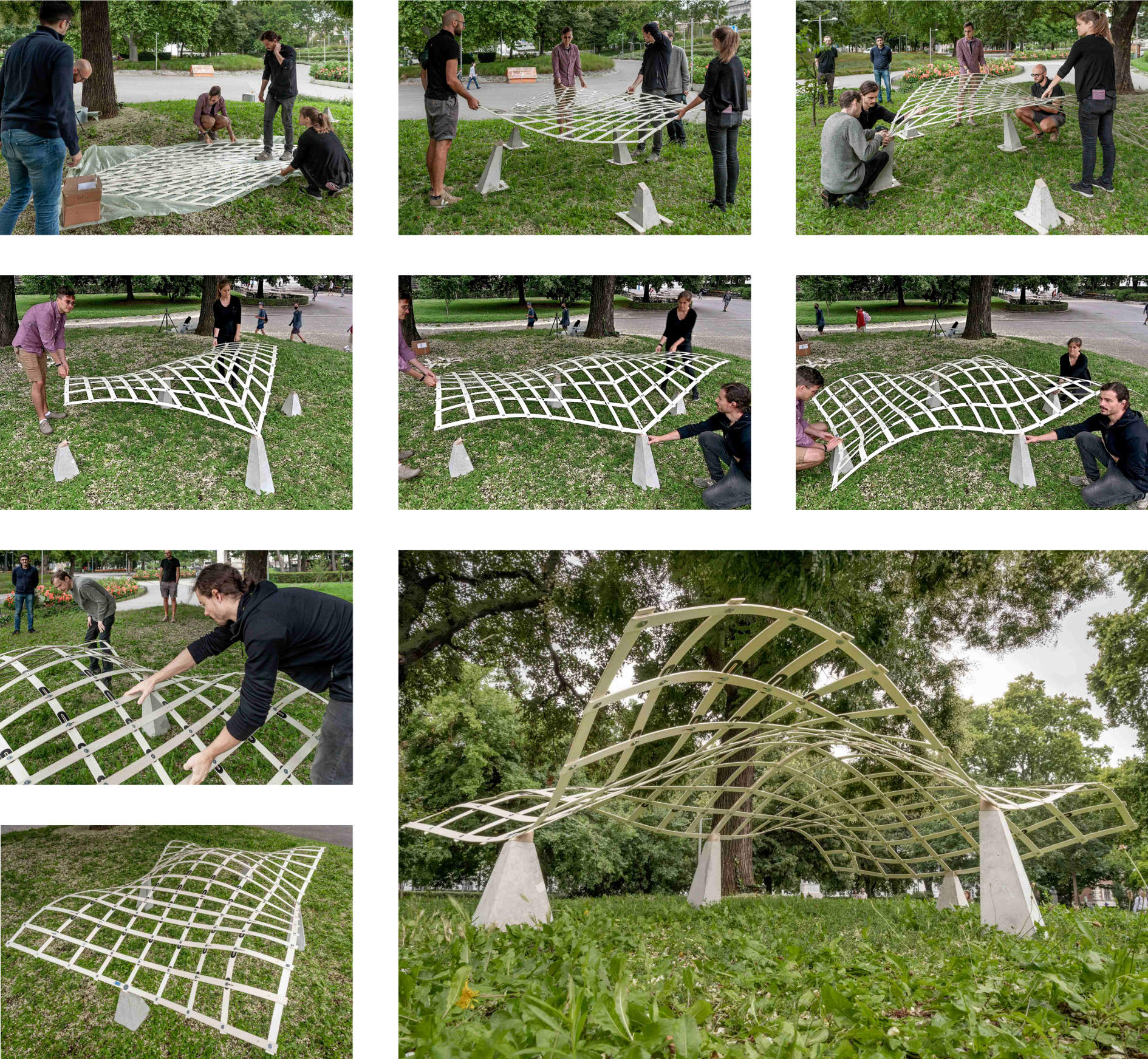}
	\caption{Several steps of the deployment of the EGG model and the deployed structure. The deployment process does not require a lot of force, and the curvature of the structure emerges naturally.}
	\label{fig:pavilion}
    \end{figure*}

	The deployment of the model was done by five people: four of them held the structure and one attached it to the supports. Some intermediate steps of the deployment process can be seen in Figure \ref{fig:pavilion}. 
	On the site, the grid was first bent and fixed to two supports, then deployed. Pre-bending the grid before deployment eliminated problems with buckling into undesired configurations, and the grid automatically deployed correctly. Deployment worked smoothly and could be carried out by a single person. Sliding of the lamellae along the notches worked partially, however, the lamellae could be pushed into the right configuration by a single person without applying much force. 
	
	\subsection{Examples from Architecture}
	
    \begin{table}[b]
        \caption{
        Model dimensions and quantitative results. We measure the root mean square error ($RMSE$) and the maximum error (max.\,$E$) between the lamellae and the design surface. }
        \begin{tabular}{rccc}
                                &  Great Court          &Lilium Tower   & Yas-Hotel  \\ [0.05cm] \midrule
            patches       & 4                     & 16            &16  \\
            size\,{[}m{]}       & 74$\times$98$\times$7 & 48$\times$48$\times$37    & 218$\times$140$\times$47 \\
            $w$ $\times$ $t$\,{[}cm{]} & 25$\times$4    & 50$\times$5       & 40$\times$5       \\
            $RMSE$\,{[}cm{]}    & 14.3                  & 24.2          & 31.8       \\
            max.\,$E$\,{[}cm{]} & 49.1                  & 176.1         & 114.1     
        \end{tabular}
        \label{tab:arch}
    \end{table}
	
\begin{figure*}
	\centering
	\includegraphics[width=\textwidth]{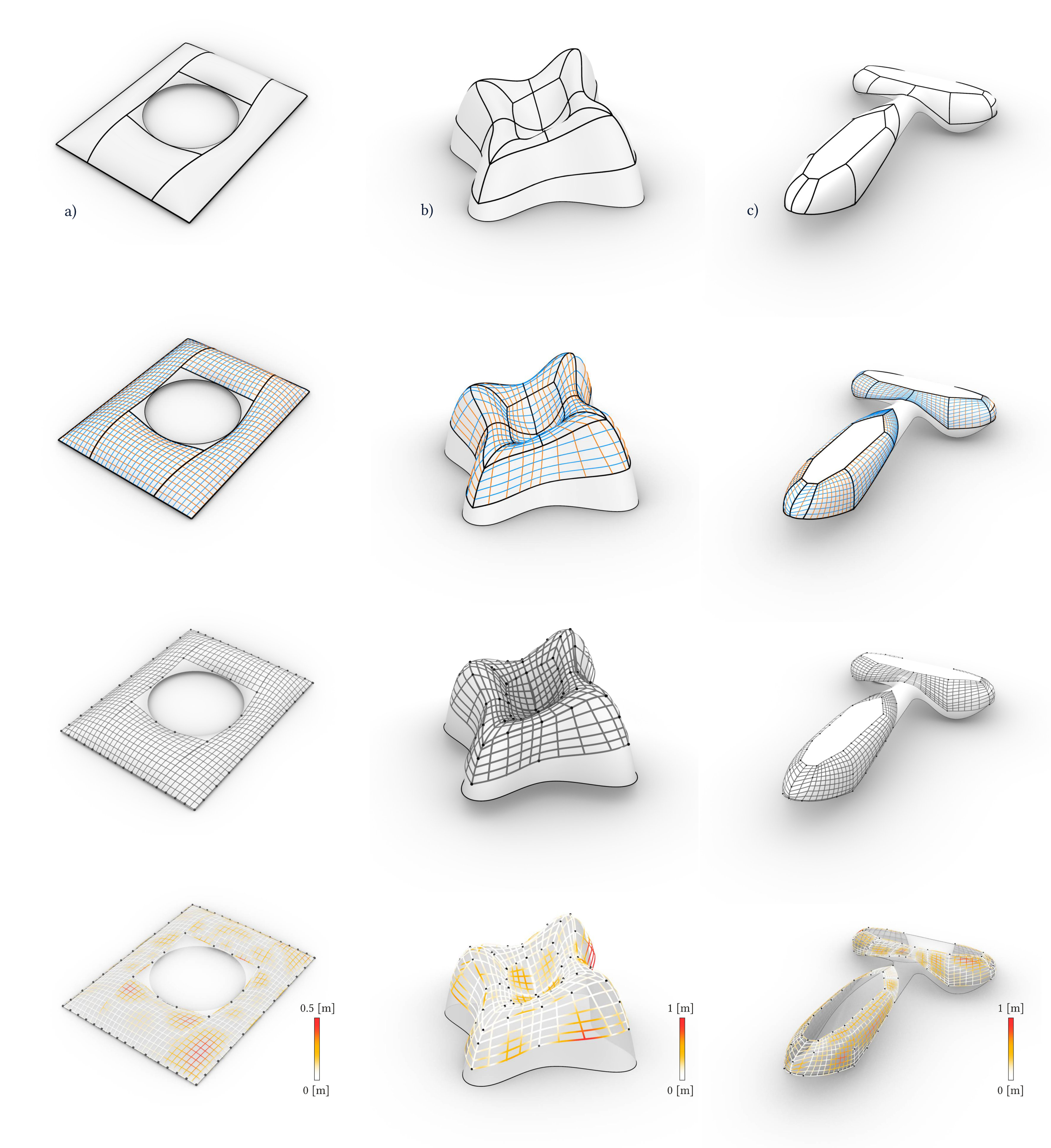} 
	\caption{Three multi-patch EGG designs at different stages. The first row shows the boundaries of the patches, the second row shows the geodesic grid members, the third row shows the equilibrium shape of the lamellae and the supports (black dots), and the last row shows a heat-map of deviations of the lamellae from the design surfaces. Columns show a) the Great Court of the British Museum, b) the tip of the Lilium tower, c) the Yas-Hotel. For numeric results, please refer to Table \ref{tab:arch}. Note that the Great Court model is not entirely symmetric, so the distribution of the error is not symmetric either.  }
	\label{fig:models_arch}
    \end{figure*}

    \step{To explore the multi-patch EGG approach, three iconic design surfaces were used:
    The Great Court of the British Museum, designed by Foster and Partners, the tip of the Lilium Tower, designed by Zaha Hadid Architects, and the Yas-Hotel, designed by Asymptote Architecture.
    The surfaces were split into patches, using the rules of Section \ref{sec:splitting_strategy}. The results are displayed in Figure \ref{fig:models_arch}. They show the design surfaces, how they are split, the geodesic grids, and the simulated shapes of the grids. The black dots indicate the position of supports. Table \ref{tab:arch} summarizes model dimensions and deviations from the design surface. 
    Please note, for designing the patches to cover the Yas-Hotel, the surface was extended and both holes were closed, all patch boundaries are shortest geodesics.
    }

	\section{Discussion and Conclusions}\label{sec:discussion}

	\step{In this paper, we investigated the idea of approximating sophisticated free-form surfaces with multiple elastic geodesic grids, to demonstrate the potential of the EGG method. 
	This was done on a theoretical and geometric level, as well as hands-on, by building a showcase model.
	We discussed the cases when the EGG approach requires a design surface to be decomposed into smaller patches and presented two basic workflows for this process.}
	
	We also presented a showcase model of some meters in size to investigate the scalability of the EGG approach. Furthermore, we analyzed some design challenges that come with these special structures, like the interaction of size and shape, smooth deployment, and the choice of material. Furthermore, we presented a simple and fast fabrication process for the model.

	\subsection{Limitations}
	
    \step{
    To validate our geometric results, we used physical simulation to compute the equilibrium shapes of the models presented in Figure \ref{fig:models_arch}. 
    This was done using the Discrete Elastic Rods model \cite{Bergou2008, Bergou2010}, with an implementation that features the simulation of self-weight \cite{Vekhter2019}.
    Please note, that this model is a standard model in computer graphics, it is fast and delivers good results, which makes it perfect for the early stages of a design process. However, to implement large-scale elastic grids, additional simulations like FEM are needed.
    
    The showcase model we presented is made of two EGG, physically connected along their common lamellae. 
    The simulated grids, displayed in Figure \ref{fig:models_arch}, are not connected along their common lamellae, which means, there is neither mutual stabilizing nor load transfer between neighboring grids. To connect the grids in the simulation and to study the influence on the overall shape is an interesting topic of its own, which we relate to future work.
    }
    
    \stepR{
    In our member strategy (Section \ref{sec:distribution}) we propagate members using the cladding functions. Please note that T-junctions of boundary members also produce a grid member, that is propagated, as in Figures \ref{fig:members} and \ref{fig:distribution}. Furthermore, grid members along a ring of patches do not need to close. In Figure \ref{fig:models_arch} for surface b) and c) this is the case only because of symmetry.
    }

    \subsection{Conclusion}
    
    \step{
    We demonstrated that the idea of multi-patch EGG is a simple and effective approach to approximate sophisticated design surfaces. The idea of covering the design surface with multiple EGG does not affect the aesthetic or structural qualities of the EGG, in fact, smaller grids are easier to handle and may therefore be more practical.
    The design surfaces from architecture were split interactively, using the rules presented in this paper, which took some hours each.
    Automating this process, incorporating certain user goals and aesthetic guidelines, is an interesting topic for future work. 
    }
    
    The comparison of the shapes of the showcase model and the small-scale prototype shows satisfying closeness. Some benchmarks for the structural performance of EGG were also presented. 
    
    \step{
    We analyzed the implication of increasing the size of the EGG, but the showcase model is only a first step in the direction of reaching architecturally relevant scales. Insights into the interaction of size and shape and the simulation results from architecture models suggest, that larger structures are feasible, especially when using high-performance, high-$\lambda$ materials.
	}

\section*{Acknowledgments}
\noindent
This research was funded by the Vienna Science and Technology Fund (WWTF ICT15-082).

\bibliographystyle{cag-num-names}
\bibliography{papers-CG,papers-additional}

\end{document}